\documentclass[journal]{IEEEtai}
\usepackage[colorlinks,urlcolor=blue,linkcolor=blue,citecolor=blue]{hyperref}
\usepackage{graphicx}
\usepackage{graphics}
\usepackage{caption}
\usepackage{lipsum}
\usepackage{orcidlink}
\usepackage[T1]{fontenc}
\usepackage[utf8]{inputenc}
\usepackage[british,UKenglish,USenglish,english,american]{babel}
\usepackage{courier}
\usepackage{authblk}
\usepackage{subcaption}
\usepackage[square,numbers]{natbib}
\setcitestyle{numbers}

\usepackage{mdframed}
\newmdenv[backgroundcolor=yellow, linecolor=yellow]{highlight}

\usepackage{booktabs,siunitx,array,threeparttable, makecell}
\usepackage{soul, color}
\usepackage{xcolor}
\usepackage{amssymb}
\usepackage{multirow}
\usepackage{makecell}
\usepackage{amsmath}
\usepackage{float}
\usepackage{listings}
\usepackage{alltt} 
\usepackage{natbib}
\usepackage{pifont}
\definecolor{darkgreen}{RGB}{0,100,0}

\newcommand{\cmark}{\textcolor{darkgreen}{\Large \ding{51}}} 
\newcommand{\xmark}{\textcolor{red}{\Large \ding{55}}}   

\usepackage{color,array}
\usepackage{framed}
\definecolor{nomadshade}{rgb}{1.0,0.94,0.94}     
\definecolor{madshade}{rgb}{0.93,1.0,0.93}       
\definecolor{nomadborder}{rgb}{0.65,0.10,0.10}   
\definecolor{madborder}{rgb}{0.10,0.45,0.10}     
\definecolor{formalshade}{rgb}{0.95,0.95,1}
\definecolor{darkblue}{rgb}{0.145, 0.118, 0.580}

\newenvironment{nomad}{%
  \MakeFramed{\advance\hsize-\width\FrameRestore}%
  \noindent\vspace{2pt}%
}
{%
  \vspace{2pt}\endMakeFramed%
}

\newenvironment{mad}{%
  \MakeFramed{\advance\hsize-\width\FrameRestore}%
  \noindent\vspace{2pt}%
}
{%
  \vspace{2pt}\endMakeFramed%
}

\definecolor{formalshade}{rgb}{0.95,0.95,1}
\definecolor{darkblue}{rgb}{0.145, 0.118, 0.580}

\newenvironment{formal}{%
  \MakeFramed{\advance\hsize-\width\FrameRestore}%
  \noindent

  \vspace{2pt}\vspace{2pt}%
}
{%
  \vspace{2pt}\endMakeFramed%
}

\usepackage{listings}
\lstdefinestyle{aiccecode}{
  basicstyle=\ttfamily\footnotesize,
  breaklines=true,
  breakatwhitespace=false,
  columns=fullflexible,
  keepspaces=true,
  showstringspaces=false,
  frame=none,
  aboveskip=0pt,
  belowskip=0pt
}

\begin{document}

\title{AICCE: AI Driven Compliance Checker Engine} 

\author{\IEEEauthorblockN{Mohammad Wali Ur Rahman\,\orcidlink{0000-0002-5009-221X}\IEEEauthorrefmark{1}, 
Martin Manuel Lopez \orcidlink{0009-0002-7652-2385}\IEEEauthorrefmark{1}, Lamia Tasnim Mim \orcidlink{0009-0005-7542-1453}\IEEEauthorrefmark{2}, Carter Farthing\IEEEauthorrefmark{3}, Julius Battle\IEEEauthorrefmark{3}, Kathryn Buckley\IEEEauthorrefmark{3},
Salim Hariri\,\orcidlink{0000-0003-3956-3401}\IEEEauthorrefmark{1} \textit{Senior Member, IEEE}}\\
\thanks{This work was supported in part by the National Science Foundation (NSF) projects under Grants 1624668, 1921485, 1303362 (Scholarship-for-Service), 2413009 and WISPER Center. Support was also provided by the 2025 Technology and Research Initiative Fund/National Security Systems Initiative administered by the University of Arizona, Office of Research and Partnerships, funded under Proposition 301, the Arizona Sales Tax for Education Act, in 2000.\\
\indent \IEEEauthorrefmark{1} Mohammad Wali Ur Rahman, Martin Manuel Lopez and Dr. Salim Hariri are with the Department of Electrical and Computer Engineering from the University of Arizona, Tucson, AZ 85721, USA. 
Email: \{mwrahman, martinmlopez, hariri\}@arizona.edu \\
\indent \IEEEauthorrefmark{2} Lamia Tasnim Mim is with Avirtek, Inc., Tucson, AZ 85712, USA. 
Email: lamiya.tasnim@avirtek.com \\
\indent \IEEEauthorrefmark{3} Carter Farthing, Julius Battle and Kathryn Buckley is with Joint Interoperability Test Command (JITC), United States Department of Defense, AZ 85613, USA. 
Email: \{carter.j.farthing.civ, julius.a.battle.civ, kathryn.v.buckley.ctr\}@mail.mil
}
}

\markboth{Journal of IEEE Transactions on Artificial Intelligence, Vol. XX, No. X, Month 2025}
{Rahman \MakeLowercase{\textit{et al.}}: AICCE: AI Driven Compliance Checker Engine}

\maketitle

\begin{abstract}
For digital infrastructure to be safe, compatible, and standards-aligned, automated communication protocol compliance verification is crucial. Nevertheless, current rule-based systems are becoming less and less effective since they are unable to identify subtle or intricate non-compliance, which attackers frequently use to establish covert communication channels in IPv6 traffic. In order to automate IPv6 compliance verification, this paper presents the Artificial Intelligence Driven Compliance Checker Engine (AICCE), a novel generative system that combines dual-architecture reasoning and retrieval-augmented generation (RAG). Specification segments pertinent to each query can be efficiently retrieved thanks to the semantic encoding of protocol standards into a high-dimensional vector space. Based on this framework, AICCE offers two complementary pipelines: (i) Explainability Mode, which uses parallel LLM agents to render decisions and settle disputes through organized discussions to improve interpretability and robustness, and (ii) Script Execution Mode, which converts clauses into Python rules that can be executed quickly for dataset-wide verification. With the debate mechanism enhancing decision reliability in complicated scenarios and the script-based pipeline lowering per-sample latency, AICCE achieves accuracy and F1-scores of up to 99\% when tested on IPv6 packet samples across sixteen cutting-edge generative models. By offering a scalable, auditable, and generalizable mechanism for identifying both routine and covert non-compliance in dynamic communication environments, our results show that AICCE overcomes the blind spots of conventional rule-based compliance checking systems.
\end{abstract}

\begin{IEEEImpStatement}
Traditional methods of compliance verification have become inadequate due to the growing complexity of communication protocol standards such as IPv6, particularly in high-assurance contexts where accuracy is of utmost importance. The methodology presented in this work combines semantic retrieval and generative multi-agent reasoning to automate the standards compliance verification process. The system provides scalable, interpretable, and model-agnostic compliance evaluations by integrating protocol requirements into a vectorized knowledge base and using retrieval-augmented generation. Decision robustness is further increased by incorporating a multi-agent discussion system, which resolves unclear or contradictory conclusions. A thorough empirical analysis using several cutting-edge generative models shows that this method consistently and accurately verifies IPv6 packet conformance. The framework's ability to semantically align real-world data with formal protocol standards represents a significant step toward dependable, AI-driven auditing tools. It is applicable not only to IPv6 but also to other domains that need dynamic, context-aware standard conformance analysis, such as defense communications, cybersecurity, and networking systems regulatory compliance.
\end{IEEEImpStatement}


\section{Introduction}
\label{sec:intro}

\IEEEPARstart{T}{hanks} to the internet's exponential growth, the proliferation of connected devices, and new application paradigms, Internet Protocol version 6 (IPv6) has become an essential part of contemporary communication systems. Central to IPv6’s adoption is its vastly expanded address space, which alleviates the exhaustion issues faced by IPv4 while enabling more flexible topological assignments~\cite{rfc8200,hinden-4291}. In addition to the expanded address space, IPv6 offers more efficient routing and opens the door for new network-layer services with its strong extension methods and simplified header structures \cite{rfc8200,deering1998rfc2460}. Large-scale service providers, mobile networks, and cutting-edge technologies like the Internet of Things (IoT) now depend on IPv6.

Despite its obvious benefits, maintaining IPv6's proper and secure operation in a variety of real-world settings is still very difficult. A complex web of interconnected Requests for Comments (RFCs) that specify different behaviors, extensions, and updates must be navigated by implementers and network operators~\cite{rfc8200,deering1998rfc2460,hinden-4291}. The complex specifications that must be standardized across heterogeneous systems frequently span hundreds of pages due to the subtle dependencies and complex requirements. Performance problems, security flaws, or incompatibilities can result from minor errors in the interpretation of important protocol clauses \cite{Dua2022,houmansadr2013,rahman2025ai}. Because IPv6 standards are dynamic and complex, and because new RFCs and clarifications are frequently released, they surpass the capabilities of conventional compliance verification techniques like static rule-checking tools or manual reviews by subject-matter experts \cite{roesch1999snort,suricata,holzmann1997spin,meier2013tamarin,sutton2007fuzzing}.

Large language models (LLMs) have recently demonstrated impressive abilities to comprehend and produce human language, opening the door to sophisticated applications in code generation, dialogue systems, and question answering~\cite{brown2020language,bommasani2021foundation, wei2022chain}. When combined with retrieval-augmented generation (RAG) techniques, these models can dynamically access external knowledge bases during inference, making their outputs more accurate and contextually grounded~\cite{lewis2020retrieval}. This method is particularly well-suited for protocol compliance verification, where the relevant sections of formal documents may be scattered across multiple RFCs. By embedding these documents into a high-dimensional vector space \cite{reimers2019sentencebert}, an LLM can retrieve highly specific clauses, parse them in context, and then generate a compliance verdict. However, applying RAG-based methods to IPv6 presents several nontrivial obstacles. The language used in protocol documents is both technical and subtle, rife with cross-references and specialized terminology. Furthermore, some clauses require careful interpretation of ambiguous text because they overlap or impose requirements that are partially contradictory.

We propose AICCE, an improved standard compliance verification framework that combines a structured debate mechanism with multi-agent generative reasoning in order to overcome these limitations. Rather than depending on a single model to determine compliance, our system coordinates the actions of several generative agents to work simultaneously. Each agent either makes an independent compliance decision or extract the important rules for compliance verification based on the context of the retrieved IPv6 specification section. A debate phase is initiated in situations where there is disagreement or uncertainty~\cite{irving2018debate,rahman2025multi}, which enables these agents to reconsider the relevant evidence. This iterative process builds a stronger consensus as agents review potential oversights and improve their decisions.  In addition to enhancing accuracy in borderline situations, this method provides an operationally transparent view of the decision-making process by identifying the exact RFC clauses and semantic interpretations that inform each decision.

We empirically validate our framework on a dataset of 1500 IPv6 packet samples, carefully curated to reflect real-world traffic diversity. We benchmark sixteen state-of-the-art generative models, assessing their performance on compliance classification metrics such as accuracy and F1-score. Our results show that the highest-performing models achieve scores above 0.99, with the debate mechanism reducing errors in challenging or ambiguous cases. Moreover, a qualitative review of the multi-agent interactions underscores how this debate mechanism uncovers hidden assumptions and offers deeper explanations for why an edge case is deemed compliant or noncompliant \cite{irving2018debate,wei2022chain}. 

In summary, the primary contributions of this paper are:
\begin{itemize}
    \item A retrieval-augmented generative AI framework specifically tailored to IPv6 compliance verification, based on a vectorized repository of official standard documents.
    \item A multi-agent debate mechanism that systematically resolves uncertain or conflicting judgments, thereby improving both classification accuracy and interpretability.
    \item A comprehensive experimental comparison of sixteen generative models, highlighting the efficacy of structured debate in high-stakes compliance contexts and offering insights into model behaviors.
\end{itemize}

While IPv6 serves as a critical focal point for our investigations, the proposed methodology is readily adaptable to other complex technical standards and regulatory environments. Accordingly, this work lays a foundation for broader applications in cybersecurity, regulatory compliance, and secure communications infrastructure. The remainder of the paper is organized as follows. Section~\ref{sec:related} provides a review of related research on network protocol verification and multi-agent reasoning. Section~\ref{sec:methodology} presents the system architecture and core algorithms that drive our framework. Section~\ref{sec:experiments} details the experimental setup, including data collection procedures and evaluation metrics. In Section~\ref{sec:results}, we analyze and interpret the empirical findings. Finally, Section~\ref{sec:conclusion} concludes the paper and outlines directions for future work.

\section{Related Articles}
\label{sec:related}

\begin{table*}[h!]
\centering
\begin{tabular}{ccccccccc}
\hline \hline
\textbf{System/Method} 
& \makecell{\textbf{Detects} \\ \textbf{Covert} \\ \textbf{Channels}}
& \makecell{\textbf{Handles} \\ \textbf{RFC} \\ \textbf{Ambiguities}}
& \makecell{\textbf{Scales} \\ \textbf{to IPv6} \\ \textbf{Extensions}}
& \makecell{\textbf{Traceable /} \\ \textbf{Explainable}}
& \textbf{Automated}
& \makecell{\textbf{Cross-} \\ \textbf{RFC} \\ \textbf{Reasoning}}
& \makecell{\textbf{Minimal}\\ \textbf{Expert Setup}}
& \makecell{\textbf{Semantic} \\ \textbf{Checking}} \\
\hline
\\
\makecell{Manual Checks \\ (Expert Review)}  
    & \cmark & \cmark & \xmark & \cmark & \xmark & \cmark & \xmark & \cmark \\ \\
Rule-Based Engines \cite{roesch1999snort, suricata} 
    & \xmark & \xmark & \xmark & \xmark & \cmark & \xmark & \cmark & \xmark \\ \\
\makecell{Formal Verification \\ (Model Checking) \\ \cite{holzmann1997spin, meier2013tamarin,sen2005smc,younes2002acceptance,david2011time,kwiatkowska2011prism4}} 
    & \cmark & \cmark & \xmark & \cmark & \cmark & \xmark & \xmark & \xmark \\ \\
Fuzz Testing \cite{sutton2007fuzzing, boofuzz, zalewski2014afl}
    & \cmark & \xmark & \cmark & \xmark & \cmark & \xmark & \cmark & \xmark \\ \\
\makecell{Single-Agent \\ System \cite{lewis2020retrieval}}
    & \xmark & \cmark & \cmark & \xmark & \cmark & \xmark & \cmark & \cmark \\ \\
\makecell{\textbf{Ours: AICCE}}
    & \cmark & \cmark & \cmark & \cmark & \cmark & \cmark & \cmark & \cmark \\
\\
\hline \hline
\end{tabular}
\caption{Qualitative Comparison of Network Protocol Validation Methods Across Key Criteria.}
\label{tab:protocol_comparison}
\end{table*}

In this section, we situate our work in the broader landscape of network protocol validation and multi-agent reasoning. We review existing methods, highlight their shortcomings in addressing the complexities of IPv6 compliance, and show how our proposed approach bridges crucial gaps.

\subsection{Network Protocol Validation}

Network protocols underpin virtually all internet communications, and verifying their correct implementation has been a focus of research for decades. Classical validation techniques rely on either extensive human oversight or purely automated testing, each with its own drawbacks. One critical area where accurate network validation is indispensable is in detecting covert communication channels within network traffic where hidden information are embedded within otherwise legitimate traffic \cite{houmansadr2013, cooper2016security}. Pattern and signature-based engines such as SNORT and Suricata \cite{roesch1999snort, suricata} are effective for known misconfigurations but frequently miss semantic deviations characteristic of covert transmissions \cite{Dua2022, rahman2025ai}.

Historically, protocol validation has often depended on domain experts meticulously comparing protocol behaviors against official specifications. Early network implementations for TCP/IP \cite{postel1981rfc791} and similar standards were checked through labor-intensive code reviews and conformance matrices, which spelled out allowed packet types, header fields, and state transitions. Over time, rule-based engines emerged to codify basic checks: for instance, verifying that IPv6 extension headers follow a specified ordering or confirming that header field values adhere to defined limits. While scalable, such static checks struggle with ambiguous or interdependent clauses spread across evolving RFCs \cite{deering1998rfc2460, hinden-4291}. Formal verification tools such as SPIN \cite{holzmann1997spin} or Tamarin \cite{meier2013tamarin} bring rigor through state-space exploration but they require substantial modeling expertise and are costly to scale to evolving protocol implementations. Complementary to exhaustive model checking, Statistical Model Checking (SMC) estimates whether a property holds by running many randomized simulations and providing results with statistical confidence \mbox{\cite{sen2005smc,younes2002acceptance}}, often scaling better when full state-space exploration is infeasible. Toolchains and approaches such as statistical checking of real-time/probabilistic systems and probabilistic verification frameworks (e.g., PRISM) have been widely used to analyze quantitative or stochastic properties \mbox{\cite{david2011time,kwiatkowska2011prism4}}. These approaches reduce up-front modeling brittleness compared to fully exhaustive exploration, but still require formalization of the system model and properties, which can be costly for evolving RFC ecosystems. Fuzzing \cite{sutton2007fuzzing} via tools like Boofuzz \cite{boofuzz} and AFL \cite{zalewski2014afl} exposes robustness defects, yet does not quantify comprehensive standards compliance.

IPv6 is a particularly challenging use case for any verification method. RFC~2460 \cite{deering1998rfc2460} was the first to outline its definition; RFC~8200 \cite{rfc8200} and IPv6 Addressing Architecture documents \cite{hinden-4291} were among the many RFCs that revised and expanded it. These papers frequently relate to one another and contain complex rules for neighbor identification, path MTU discovery, address scopes, extension header chaining, and more. As a result, it is easy for implementers to miss subtle interactions between several standard parts. Until complicated, real-world circumstances, such as tunneling or traffic in heavily meshed topologies, disclose hidden assumptions, errors or ignored needs may go undetected \cite{huston2020measuring}. Although manual validation, mathematically grounded formal verification, and fuzz testing each address specific facets, they fall short of comprehensive, end-to-end assurance of clause-faithful IPv6 compliance at scale. Manual reviews do not scale, formal methods are costly and brittle to evolving implementations, and fuzzing reveals robustness defects rather than systematic conformance to interdependent RFC clauses. Moreover, none of these approaches reliably captures the semantic reasoning required to interpret cross-referenced, context-dependent requirements in the IPv6 standards.

\subsection{Multi-Agent Reasoning}

The field of multi-agent AI has grown rapidly in tandem with improvements in protocol validation. Modern applications take advantage of the combined intelligence of several independent agents, each of which offers unique insights or partial answers to complex problems.

In multiagent AI systems, multiple intelligent agents collaborate (or sometimes compete) to accomplish tasks whose complexity may surpass the capability of a single agent. Each agent acts on local observations yet may also exchange information or learned parameters to enhance collective outcomes \cite{wooldridge2009introduction}. This paradigm has proven effective in highly distributed settings such as federated learning \cite{konevcny2016federated}, where data is scattered across different nodes or devices and must be processed without compromising privacy. Multiagent reinforcement learning (RL-MAS) extends single-agent RL principles to shared environments with multiple decision-makers \cite{busoniu2008comprehensive}. Agents iteratively refine their policies via trial and error, adapting to evolving conditions and the actions of other agents. 

Swarm intelligence approaches, such as particle swarm optimization \cite{kennedy1995particle} and ant colony optimization \cite{dorigo2019ant}, exemplify how decentralized groups can coordinate to solve optimization problems by exploring a wide range of solutions in parallel. Negotiation-based models, meanwhile, allocate resources and resolve conflicts through formal agreement mechanisms, often informed by game-theoretic insights \cite{kraus1997negotiation, fatima2009analysis}. Among these, actor-critic methods \cite{konda2000actor} have gained attention because they couple policy-based and value-based learning in a single framework, allowing for dynamic adjustments based on ongoing evaluative feedback. This iterative refinement process is especially valuable in scenarios that demand adaptive, high-precision decision-making.

While the literature on multi-agent systems (MAS) is extensive, applications to network protocol compliance remain limited. Typical MAS formulations presume well-defined environments and explicit payoffs, whereas IPv6 conformance requires interpreting interlinked, sometimes ambiguous RFC clauses. Domain-agnostic or heuristic debate structures thus risk shallow, ungrounded arguments unless anchored in standards text; without robust retrieval, agent discussions can miss critical evidence and yield spurious conclusions.

To address this gap, prior advances in retrieval-augmented generation (RAG) \cite{lewis2020retrieval} and structured debate \cite{irving2018debate} motivate clause-grounded, multi-agent reasoning for compliance. Embedding IPv6 RFCs enables agents to cite explicit passages and contest interpretations (e.g., extension-header sequencing), promoting consensus that is both auditable and accurate. In contrast, black-box single-agent methods obscure evidentiary chains, undermining correctness and interpretability in high-stakes networking. In sum, whereas manual checks, fuzzing, and formal proofs miss cross-document semantics, and generic MAS lacks domain grounding, RAG-backed debate directly targets clause-faithful verification; subsequent sections detail the architecture and empirical benchmarks. Table~\mbox{\ref{tab:protocol_comparison}} compares representative protocol-validation approaches for IPv6. \textit{Automated} denotes whether the pipeline runs end-to-end at runtime once its required artifacts (e.g., rules, formal models, or specification encodings) are in place, without interactive human involvement during checking. \textit{Minimal Expert Setup} captures the up-front and maintenance effort to create and update those artifacts as RFCs evolve; thus, model checking is automated at runtime but typically has high expert setup cost.

\begin{figure*}[!t]
    \centering
    \captionsetup{justification=centering}
    \includegraphics[width=\textwidth]{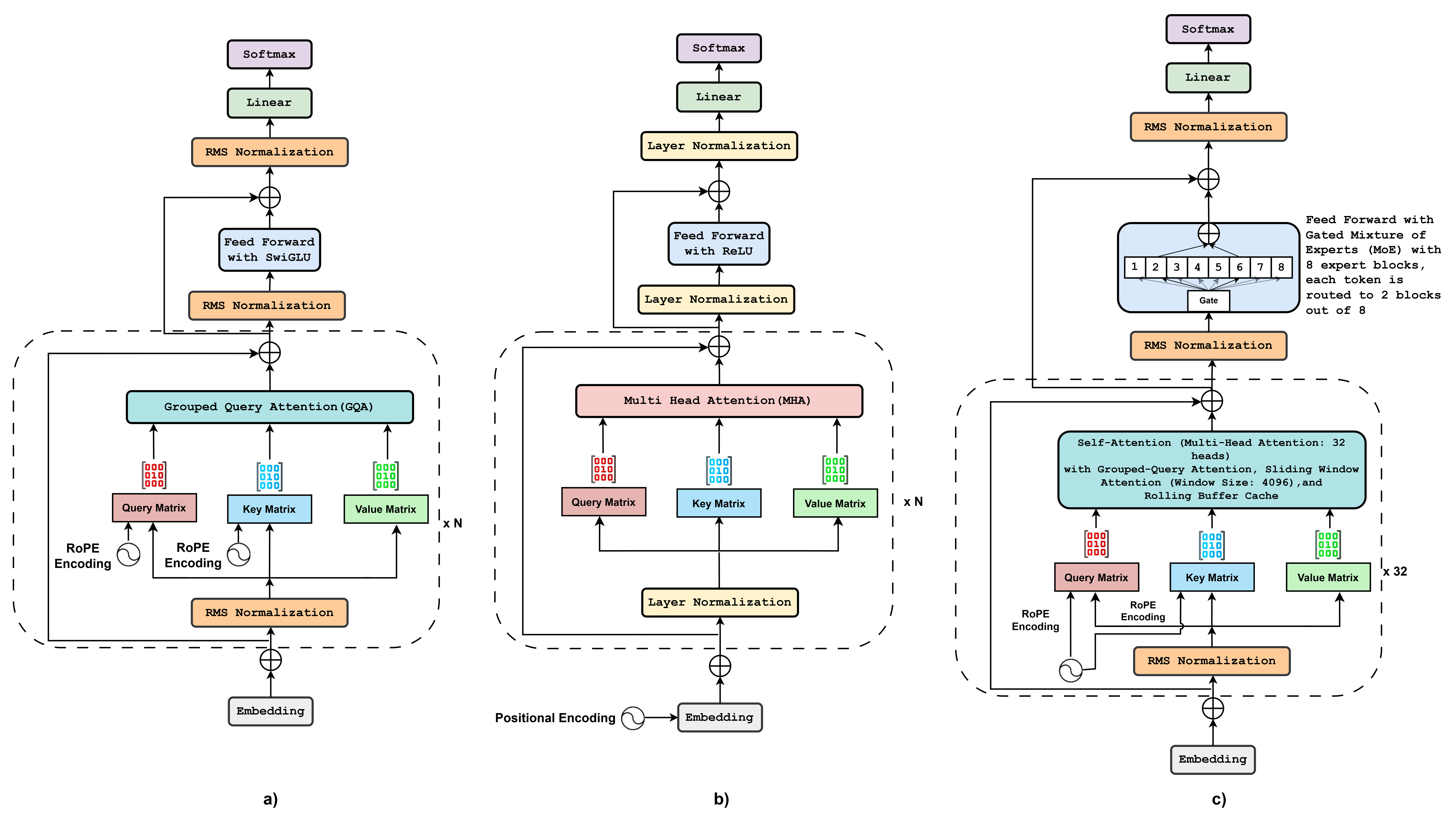}
    \caption{Basic Architecture Diagram of a) Llama, b) GPT and c) Mixtral 8x7b }
    \label{fig:llm_archi}
\end{figure*}

\section{Methodology}
\label{sec:methodology}

\begin{figure*}[!t]
    \centering
    \captionsetup{justification=centering}
    \includegraphics[width=\textwidth]{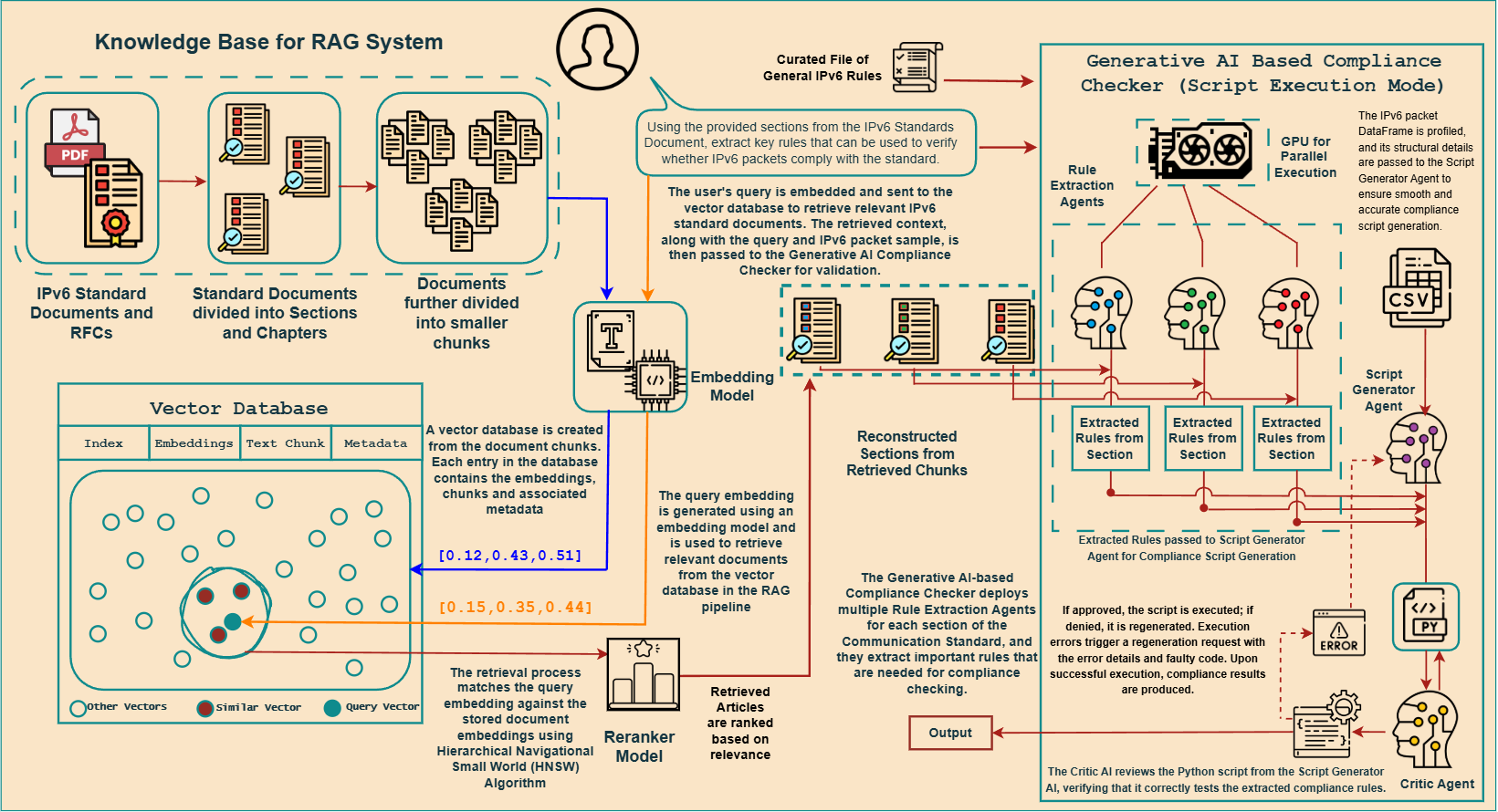}
    \caption{AICCE Framework in Script Execution Mode}
    \label{fig:framework2}
\end{figure*}

This section presents the complete methodology behind our generative AI–based IPv6 protocol compliance framework, which is implemented in two distinct architectures: the\textbf{ Explainability Mode} for deep interpretability and thorough clause-level compliance verification, and the \textbf{Script Execution Mode} for low-latency execution. Both pipelines rely on semantic document embeddings, vector-based retrieval, and dynamic standard interpretation. Their complementary strengths where one favors depth and auditability, the other speed and automation, enable flexible deployment across different operational constraints. The overall system architecture is illustrated in Figure~\ref{fig:framework1} and Figure~\ref{fig:framework2}, and their components are elaborated below.

\subsection{Semantic Encoding of Protocol Standards}
\label{subsec:semantic_encoding}

At the foundation of both proposed architectures lies the semantic encoding of protocol standards, which transforms unstructured textual specifications into structured, machine-readable representations suitable for retrieval and compliance reasoning. The objective is to map natural language descriptions of IPv6 RFC clauses into high-dimensional embedding vectors that preserve semantic meaning and contextual dependencies. This enables efficient similarity computation between packet-level compliance queries and relevant RFC passages, thereby forming the basis of the retrieval-augmented compliance framework \cite{reimers2019sentencebert}.

\subsubsection{Document Preprocessing and Chunking}

All standard documents, including IPv6 RFCs and their extensions, are first preprocessed to extract textual content and remove extraneous formatting artifacts. The text is then segmented into semantically coherent units, or \textit{chunks}, each limited to a maximum of 512 tokens in order to align with transformer model input constraints while retaining local context. Formally, for a given RFC document \(D\), we obtain a sequence of contiguous chunks:
\begin{equation}
D = \{C_1, C_2, \dots, C_N\},
\end{equation}
where each \(C_i\) denotes the \(i^{th}\) chunk and \(N\) is the total number of extracted segments. This chunking strategy preserves semantic integrity while ensuring tractable embedding computation.

\subsubsection{Transformer-Based Embeddings}

Each chunk \(C_i\) is projected into a high-dimensional semantic space using a sentence transformer, a pretrained model optimized for semantic search tasks. The embedding process can be expressed as:
\begin{equation}
\mathbf{v}_i = \text{TransformerEncoder}(C_i),
\end{equation}
where \(\mathbf{v}_i \in \mathbb{R}^{768}\) is the resulting embedding vector for chunk \(C_i\). The encoder leverages multi-head self-attention and pooling operations (e.g., mean pooling) to integrate contextual dependencies across sentences. 

At each transformer layer \(l\), the hidden state representation \(\mathbf{H}^{(l)}\) is computed via self-attention:
\begin{equation}
\mathbf{H}^{(l)} = \text{SelfAttention}(\mathbf{H}^{(l-1)}),
\end{equation}
where the self-attention mechanism itself is defined as:
\begin{equation}
\text{SelfAttention}(Q, K, V) = \text{softmax}\left(\frac{QK^\top}{\sqrt{d_k}}\right)V,
\end{equation}
with query, key, and value matrices \(Q\), \(K\), and \(V\), and \(d_k\) representing the dimensionality of the queries and keys. The final embedding \(\mathbf{v}_i\) is taken from the output of the last transformer layer after pooling.

\subsubsection{Semantic Similarity}

The embeddings are designed such that semantic similarity between two chunks corresponds closely to their vector similarity. This is measured using cosine similarity:
\begin{equation}
\text{sim}(\mathbf{v}_i, \mathbf{v}_j) = \frac{\mathbf{v}_i \cdot \mathbf{v}_j}{\|\mathbf{v}_i\|\|\mathbf{v}_j\|},
\end{equation}
where a higher similarity score reflects stronger semantic relatedness between chunks \(C_i\) and \(C_j\). 

These embeddings, once generated, are indexed into a vector database for efficient retrieval, as discussed in Section~\ref{subsec:vector_database}. The semantic encoding step thus provides the critical bridge between natural-language protocol standards and automated compliance verification, enabling both architectures to operate on a unified semantic representation of IPv6 specifications.

\subsection{Vector Database and Semantic Retrieval}
\label{subsec:vector_database}

To enable efficient semantic access to IPv6 protocol specifications, the high-dimensional embeddings produced during semantic encoding are organized and indexed within a persistent vector database. Given the computational intractability of linear similarity search in large embedding spaces, approximate nearest neighbor (ANN) indexing is employed to achieve both scalability and low-latency retrieval. Specifically, we adopt the Hierarchical Navigable Small World (HNSW) graph \cite{malkov2018efficient}, a graph-based ANN structure that is widely recognized for its robustness in high-dimensional search tasks.

\subsubsection{HNSW Graph Indexing}

Each chunk embedding derived from the RFC corpus is stored as a node in the HNSW graph. Edges between nodes are induced by cosine distance, so semantically proximate embeddings form densely connected local neighborhoods. Formally, let
\[
V = \{\mathbf{v}_1, \mathbf{v}_2, \dots, \mathbf{v}_N\}, \quad \mathbf{v}_i \in \mathbb{R}^{768},
\]
denote the set of all stored embeddings, and let \mbox{\(d(\cdot,\cdot)\)} be cosine distance. For a given node \mbox{\(\mathbf{v}_i\)}, let \mbox{\(\mathbf{v}_{(k)}\)} denote its \mbox{\(k\)}-th nearest neighbor in \mbox{\(V\setminus\{\mathbf{v}_i\}\)} under \mbox{\(d(\cdot,\cdot)\)}. We define the \mbox{\(k\)}-radius neighborhood of \mbox{\(\mathbf{v}_i\) as}
\begin{equation}
U_k(\mathbf{v}_i)=\left\{\mathbf{v}_j \in V\setminus\{\mathbf{v}_i\}\ \middle|\ d(\mathbf{v}_i,\mathbf{v}_j)\le d(\mathbf{v}_i,\mathbf{v}_{(k)})\right\},
\label{eq:hnsw_neighborhood}
\end{equation}
which captures all embeddings at least as close to \mbox{\(\mathbf{v}_i\)} as its \mbox{\(k\)}-th nearest neighbor (i.e., a distance-thresholded local neighborhood). In HNSW, nodes are inserted probabilistically into multiple hierarchical layers: higher layers provide sparse long-range links that act as navigational shortcuts, while lower layers preserve dense local connectivity. This hierarchical structure enables efficient approximate nearest-neighbor search with low query latency in high-dimensional embedding spaces.

\begin{figure}[!h]
    \centering
    \captionsetup{justification=centering}
    \includegraphics[width=0.9\columnwidth]{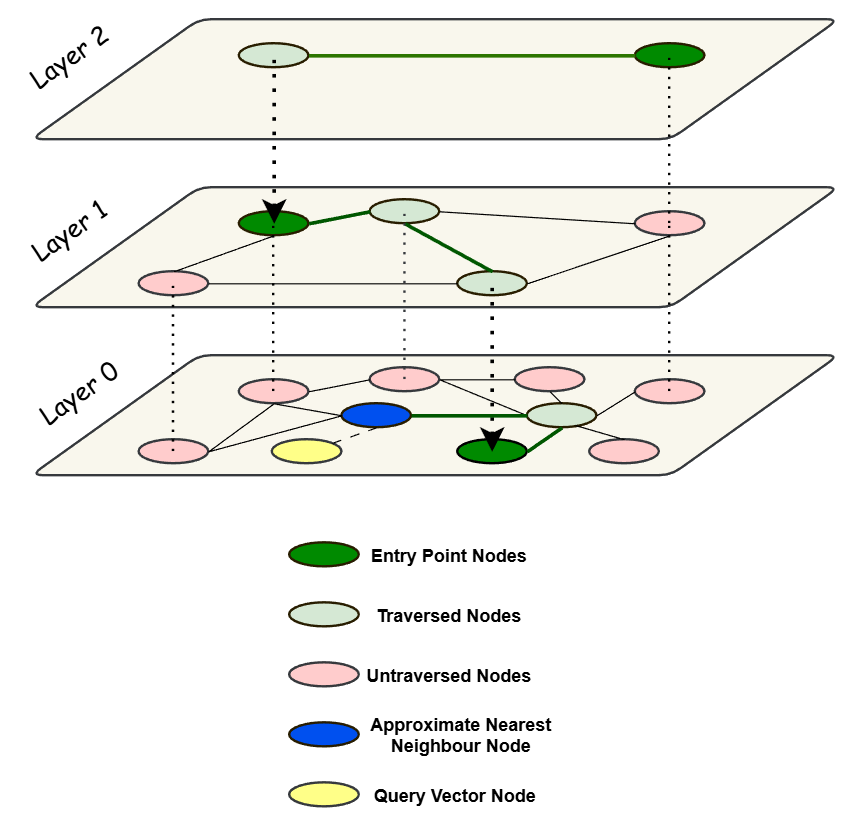}
    \caption{Visualization of ANN search with the Hierarchical Navigable Small World (HNSW) algorithm.}
    \label{fig:hnsw}
\end{figure}

\begin{figure*}[!t]
    \centering
    \captionsetup{justification=centering}
    \includegraphics[width=\textwidth]{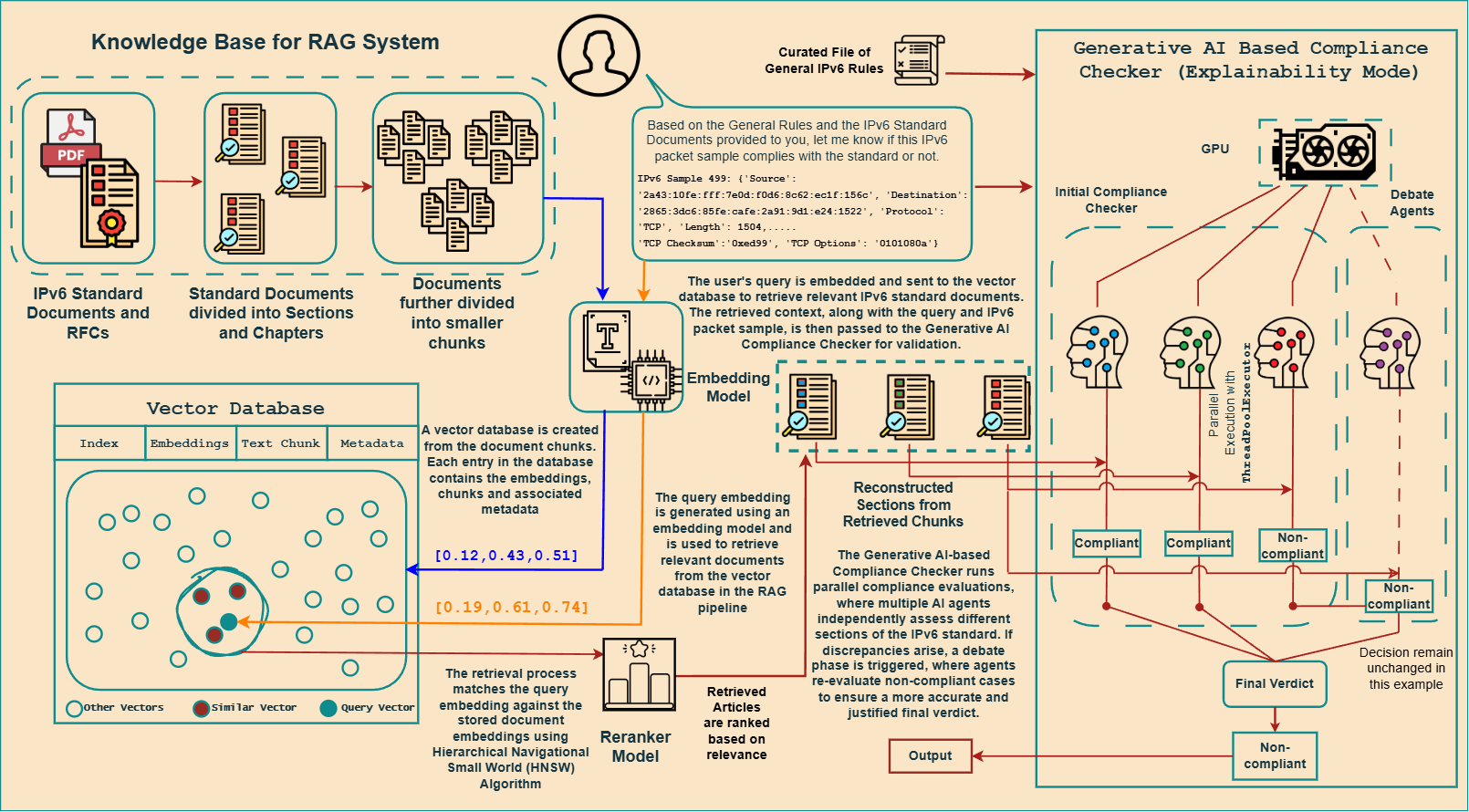}
    \caption{AICCE Framework in Explainability Mode}
    \label{fig:framework1}
\end{figure*}

\subsubsection{Query-Time Retrieval}

At query time, a compliance-related input, such as a packet description or a field-specific question, is encoded into an embedding vector \(\mathbf{q} \in \mathbb{R}^{768}\) using the same sentence transformer employed during document encoding. The retrieval objective is to identify the top-\(k\) nearest neighbors to \(\mathbf{q}\) according to cosine similarity:
\begin{equation}
R(\mathbf{q}) = \arg\max_{\mathbf{v}_i \in V, \, |R|=k} \, \text{sim}(\mathbf{q}, \mathbf{v}_i),
\end{equation}
where similarity is defined as:
\begin{equation}
\text{sim}(\mathbf{q}, \mathbf{v}_i) = \frac{\mathbf{q} \cdot \mathbf{v}_i}{\|\mathbf{q}\| \|\mathbf{v}_i\|}.
\end{equation}

To ensure retrieval precision, we apply a similarity threshold \(\tau\), such that only embeddings sufficiently close to the query are retained:
\begin{equation}
R_{\tau}(\mathbf{q}) = \{ \mathbf{v}_i \in R(\mathbf{q}) \mid \text{sim}(\mathbf{q}, \mathbf{v}_i) \geq \tau \}.
\end{equation}
This empirically chosen threshold was found to effectively suppress spurious matches while retaining all semantically relevant IPv6 clauses. Figure \ref{fig:hnsw} illustrates the process of ANN search combined with Hierarchical Navigational Small World (HNSW) Algorithm.

\subsubsection{Reconstructed Sections}

Since individual chunks often represent only partial clauses of the standard, retrieval results from the same source document are concatenated to form \textit{reconstructed sections}. These reconstructed sections provide semantically coherent context windows that preserve both local meaning and broader document continuity. They serve as the primary knowledge units for downstream compliance reasoning, ensuring that generative agents evaluate packet validity against consistent and contextually rich evidence. Compliance classification assumes the full packet record is available; if fields are missing, some rules become unverifiable and the same accuracy cannot be guaranteed. RFC chunking does not weaken reasoning because retrieved chunks are merged into reconstructed sections, and the framework retrieves all relevant sections for each query by scaling agent deployment to the number of sections returned, ensuring clause-complete (not fragmentary) context. Compared to traditional keyword-based search, semantic retrieval via HNSW indexing provides significant advantages. Keyword systems fail when terminology varies (e.g., ``payload length'' vs. ``size of data field''), while embedding-based retrieval aligns semantically equivalent but lexically distinct phrases. Moreover, the HNSW structure ensures efficient scalability, reducing query latency to sub-second levels even for large-scale document collections.

\subsubsection{Model Families and Architectural Diversity}
\label{subsec:model_families}

The reconstructed sections produced by semantic retrieval serve as the authoritative context for all downstream reasoning. Each section aggregates contiguous, clause-coherent text from a single RFC, preserving local semantics and cross-references. We pass these sections verbatim, along with lightweight metadata identifying the source RFC and clause boundaries to the generative agents. We evaluate distinct large language model (LLM) families to probe how architectural choices affect retrieval grounding, clause interpretation, and downstream agent behavior. Specifically, we consider OpenAI’s GPT series, Meta’s Llama models, Anthropic’s Claude models, Google’s Gemini family, Alibaba’s Qwen series, and open-source families including Mistral AI’s Mixtral and Mistral models and the Gemma models developed through the Google–NVIDIA collaboration. As illustrated in Fig.~\ref{fig:llm_archi}, Llama variants employ grouped-query attention (GQA), RMS normalization, and rotary positional embeddings (RoPE); GPT-family models use multi-head attention (MHA), layer normalization, and learned/rotary positional schemes; and Mixtral augments transformer layers with a gated mixture-of-experts (MoE) router and grouped/sliding-window attention. This architectural diversity materially impacts reasoning style, long-context handling, and error modes. In the experiments that follow, we therefore adopt single-model ensembles, instantiating all agents for a given run from the same backbone, to enable controlled, within-model comparisons of section-level judgments and debate dynamics.

\subsection{Dual-Architecture Generative Compliance Evaluation}
\label{subsec:dual_architectures}

To operationalize the semantic retrieval outputs for IPv6 compliance verification, we designed two complementary generative architectures. Both consume reconstructed protocol sections from the vector database and apply large language models (LLMs) to evaluate compliance, but they differ fundamentally in how the packet evaluation is performed. In the first architecture, each LLM agent is assigned to a specific protocol section and independently assesses individual IPv6 packets with respect to that section, issuing a section-specific verdict and rationale. In contrast, the second architecture does not expose LLMs to packet data directly; instead, agents extract executable compliance rules from their assigned sections, which are subsequently composed into a unified Python script. This script is then applied to the dataset for high-speed, rule-based verification. Together, these two approaches highlight complementary strengths, section-level interpretability and deliberation in the first case, and dataset-wide automation and latency reduction in the second, thereby mapping out the trade-off space between transparency and efficiency in generative compliance systems.

\subsubsection{Architecture A (Explainability Mode)}
\label{subsec:archA}

The retrieved protocol specification segments obtained through semantic retrieval form the contextual foundation for our generative multi-agent compliance verification system. In this approach, each protocol compliance query initiates the parallel deployment of multiple generative AI agents, each tasked with independently evaluating the provided semantic context to determine IPv6 packet compliance. Rather than integrating heterogeneous generative models within a single ensemble, our system is organized as \emph{single-model} ensembles: each experiment instantiates all agents from the same state-of-the-art LLM, enabling a clean, within-architecture assessment of interpretive strengths, biases, and consistency across sections. Figure \ref{fig:framework1} illustrates the framework for AICCE in Explainability Mode.


For a given packet \(Q\) and reconstructed section \(C_i\), a dedicated agent \(A_i\) receives the section text, a curated file of general IPv6 rules \(R_g\), and the full packet features. Each agent returns a binary verdict with rationale in response to: \emph{``Does the packet satisfy all required compliance properties under this section?''} Formally, the initial assessment is
\begin{equation}
\label{eq:archA_initial}
V_i^{(0)} = \mathrm{LLM}_i(Q, C_i, R_g),
\end{equation}
where \(\mathrm{LLM}_i\) denotes an instance of the chosen backbone model. Inference settings are held fixed for comparability and all agents for a packet run concurrently via a thread-based executor.

If all section agents return \texttt{Yes}, the packet is accepted; otherwise, a structured debate is triggered to resolve conflicts and surface clause-level justifications. At debate iteration \(t\!\ge\!1\), each agent revises its verdict after being shown peer rationales from round \(t\!-\!1\):
\begin{equation}
\label{eq:archA_debate}
V_i^{(t)} = \mathrm{LLM}_i\!\big(Q, C_i, R_g, \{V_j^{(t-1)}\}_{j\neq i}\big).
\end{equation}
Debate rounds for each LLM execute in parallel and are capped at five to bound latency; convergence to unanimous compliance leads to acceptance, else the packet is declared non-compliant. Inspired by recent advancements in multi-agent generative AI research \cite{rahman2025multi}, this design yields clause-level interpretability and systematically mitigates single-agent brittleness by exposing and adjudicating disagreements through informed counterargumentation.

\subsubsection{Architecture B: Script Execution Mode}
\label{subsec:archB}

The second architecture pursues a complementary objective: minimize decision latency while retaining auditability through explicit, executable checks. It does so by translating textual clauses into Python-level rules and running them directly over the dataset.

Each reconstructed section is passed to a \emph{Section Rule Extraction} agent that reads the section text and \(R_g\), cross-checks available fields in the DataFrame schema, and emits only actionable rules involving present columns. Rules are expressed as single-line predicates. A simplified example could be:
\begin{equation}
\label{eq:rules_examples}
\begin{aligned}
&\texttt{packet["Version"] == 6} \\
&\texttt{packet["HopLimit"] < 256}
\end{aligned}
\end{equation}

A \emph{Script Generator} aggregates the rules into a compliance checking function that computes a Boolean column per rule and a final \texttt{Compliant} column (logical \textsc{and} over all rules). A \emph{Critic} reviews the script against the rules and schema; if deficiencies are found, a refinement loop ensues:
\begin{equation}
\label{eq:script_refine}
S^{(t+1)} = \mathrm{ScriptGen}\!\big(R,\, \mathrm{Critic}(S^{(t)}, R)\big),
\end{equation}
which repeats until the critic returns \texttt{APPROVED} or a turn limit is reached. If runtime errors occur, the traceback is fed back to Script Generator for targeted repair. Once approved, the script executes over the full dataset, yielding a verdict per row plus rule-level diagnostics. This pipeline attains sub-second per-packet latency (approximately \(0.8\)s on average), offering substantial throughput gains versus debate.

\subsubsection{Parallelism, Coverage, and Aggregation}
\label{subsec:parallel_agg}

In both architectures, the number of concurrently instantiated units adapts to the number of reconstructed sections returned by retrieval, ensuring each clause is evaluated in context. Execution is parallelized using a thread pool to reduce wall-clock latency. Architecture~A aggregates section-level verdicts after debate, requiring unanimity for compliance; Architecture~B aggregates rule columns via logical conjunction to form row-level labels. The former yields rich rationales and cross-checked judgments, while the latter provides a sparse, auditable rule matrix suited to high-throughput screening.

\subsubsection{Comparison of architectures}
\label{subsec:arch_compare}

Architecture~A foregrounds interpretability, thorough section-by-section reasoning, and explicit adjudication of disagreement; Architecture~B prioritizes speed and deterministic execution via dataset-aware rules. Table~\ref{tab:architecture_comparison} summarizes the principal trade-offs.

\begin{table}[h]
\centering
\caption{Comparison of Framework Architectures}
\label{tab:architecture_comparison}
\begin{tabular}{|l|c|c|}
\hline
\textbf{Aspect} & \textbf{Architecture A} & \textbf{Architecture B} \\
\hline
Primary Focus & Interpretability, coverage & Speed, automation \\
Multi-agent Usage & per section \& per packet & per section \\
Debate Mechanism & Yes  & Yes\\
Granularity & Section-level judgments & Rule-level execution \\
Output & Rationale + verdict & Rule matrix + verdict \\
Exec Speed & $\sim 2$\,s / packet & $\sim 0.8$\,s / packet\\
\hline
\end{tabular}
\end{table}

\section{Experiments}
\label{sec:experiments}

This section describes the experimental evaluation conducted to assess the performance of the proposed IPv6 compliance verification framework. The experimental protocol begins with an Exploratory Data Analysis (EDA) to clearly characterize the dataset, including its origins, data integrity considerations, and non-compliance anomalies introduced for comprehensive testing.

\subsection{Exploratory Data Analysis}
\label{subsec:eda}

The dataset utilized for evaluating IPv6 compliance detection was sourced from the Center for Applied Internet Data Analysis (CAIDA). Specifically, the Day 1 IPv6 trace dataset was selected due to its high data integrity, minimal corruption rate, and representative coverage of real-world IPv6 traffic. CAIDA’s datasets are widely recognized and employed in network research due to their comprehensive and clean capture of genuine Internet traffic patterns, thereby ensuring realistic and reproducible experimental conditions. 

Initially, we randomly sampled 1500 packets from the Day 1 dataset, and these samples were all considered standards-compliant upon selection.

To rigorously evaluate the framework's ability to detect protocol anomalies, a subset of non-compliant samples was generated through intentional violations. Specifically, 300 randomly selected compliant samples were modified by applying one to three distinct violations, thereby rendering these packets non-compliant according to IPv6 standards. Due to the randomized violation selection, the total number of violations exceeds the number of non-compliant samples, leading to richer testing scenarios.

Table~\ref{tab:difficulty_distribution} summarizes the frequency of each introduced violation across the 300 non-compliant samples. Although exactly 300 samples were modified, the total number of individual violations introduced was 578, given that multiple violations were often applied per sample.

Each violation was further categorized by detection difficulty into three classes: \textbf{Trivial}, \textbf{Moderate}, and \textbf{Challenging}, based on how easily an anomaly could be identified or inferred through simple rule-based checks versus requiring sophisticated semantic reasoning. Table~\ref{tab:difficulty_distribution} provides this categorization and the frequency of each violation type per category. The details of each violation type has been discussed in \textbf{Appendix C}.

\begin{table}[htbp]
\centering
\caption{Violation Types by Detection Difficulty}
\label{tab:difficulty_distribution}
\begin{tabular}{lcc}
\toprule
\textbf{Violation Type} & \textbf{Difficulty} & \textbf{Occurrences}\\ 
\midrule
IPv6 Version Violation         & Trivial      & 52 \\ 
Length Field Violation         & Moderate     & 54 \\[2pt]
IPv6 Address Violation         & Trivial      & 54 \\[2pt]
Hop Limit Violation            & Moderate     & 52 \\[2pt]
TCP/UDP/ICMPv6 Overlap                & Challenging  & 49 \\[2pt]
Flow Label Violation           & Moderate     & 54 \\[2pt]
DSCP Field Violation           & Moderate     & 63 \\[2pt]
ECN Field Violation            & Moderate     & 49 \\[2pt]
Protocol Mismatch Violation    & Challenging  & 52 \\[2pt]
Next Header Field Violation    & Challenging  & 46 \\[2pt]
ICMPv6 Code Violation          & Moderate     & 53 \\[2pt]
\bottomrule
\end{tabular}
\end{table}

This EDA clarifies the structure and challenges posed by the constructed dataset. The deliberate injection of diverse and multiple violations per sample generates a comprehensive set of scenarios that thoroughly stress-test the robustness and accuracy of the proposed compliance detection system. By explicitly categorizing these violations, the dataset provides both straightforward and nuanced challenges for system evaluation, thus laying a solid foundation for experimental validation.

\subsection{Experiment Settings}
\label{subsec:experiment_settings}

All evaluations were executed under a single, deterministic pipeline to ensure
comparability across large–language–model (LLM) ensembles and to facilitate
full reproducibility. Identical code, prompts, and hyper-parameter values were used for every LLM provider.

\subsubsection{Vector–database construction}
Official IPv6 RFCs and relevant addenda were downloaded in PDF form and
automatically segmented into sentence-level chunks with an upper bound of
$512$ tokens per chunk.  Each chunk was embedded with a 32-layer sentence-transformer optimised for semantic search \cite{reimers2019sentencebert}.  The resulting
$d=768$-dimensional vectors were persisted in a ChromaDB collection, indexed with an HNSW graph (\texttt{hnsw:space = cosine}) for sub-second approximate-nearest-neighbour lookup \cite{malkov2018efficient}.  Every record stores (i)~the raw text, (ii)~its embedding, and (iii)~a single JSON metadata key \texttt{document\_name}, enabling rapid source tracing during error analysis.

\subsubsection{Semantic retrieval}
At run time an ad-hoc query embedding is produced with the same transformer and
matched against at most $k=50$ neighbours.  A similarity threshold of
$\tau = 0.60$ is applied to filter low-relevance matches; this value was chosen
empirically as the smallest threshold that suppressed obvious false positives
while retaining all ground-truth IPv6 clauses in a validation sweep.
Retrieved chunks belonging to the same RFC are concatenated into a
\emph{reconstructed section}. Ranking is solely by HNSW cosine distance and a lightweight metadata-based reordering is performed in which chunks are grouped and prioritized by their \texttt{document\_name} field. This ensures that related segments from the same RFC are surfaced together, preserving clause continuity and contextual integrity.

\subsubsection{Packet-level compliance agents (Architecture A)}
For Architecture~A (\emph{Explainability Mode}), every reconstructed section spawns
an independent compliance agent. Each agent is a single LLM call that receives:
(i)~the reconstructed section, (ii)~the global IPv6 general-rules file, and
(iii)~the packet under inspection encoded as a JSON dictionary.
All LLMs are executed with fixed inference settings
(\texttt{temperature} = 0.7, \texttt{top\_p} = 1.0, provider-default
\texttt{top\_k}).  Agents return an explicit \texttt{Yes}/\texttt{No}/\texttt{Maybe} verdict
plus a one-sentence rationale.  All section agents for a given packet are run
concurrently via Python’s \texttt{ThreadPoolExecutor} (\texttt{max\_workers}=8).

\subsubsection{Multi-agent Debate (Architecture A)}
When an agent expresses uncertainty in its initial verdict, or when at least two
section-specific agents disagree, a structured debate phase is triggered. For
each disputed section, a new ``Debate Agent’’, implemented as a single LLM call
with the same inference hyper-parameters, is prompted with the uncertain or
conflicting rationales, together with the original section, the general rules,
and the packet. The debate agent must then produce a refreshed
\texttt{Yes}/\texttt{No} decision, prefixed accordingly, followed by a concise
justification. Debate rounds are capped at five at the LLM level; since each packet is checked across multiple sections, multiple debates may be invoked per packet. A packet is labelled \emph{compliant} only if \emph{all} section agents agree after the final debate round; otherwise it is declared \emph{non-compliant}. In practice, most debates converge in $\leq 2$ rounds, though the framework allows up to five rounds per disputed section. 

\subsubsection{Rule Extraction Agents (Architecture B)}
For Architecture B, each reconstructed section is processed by a
dedicated \emph{Rule Extraction Agent}. These agents operate independently and
map textual clauses from the RFC sections and global IPv6 compliance rules
onto executable logical conditions. To maintain alignment with the dataset,
rules that reference fields absent from the packet schema are automatically
discarded. Each agent is a single LLM call configured with fixed inference
settings (\texttt{temperature} = 0.7, \texttt{top\_p} = 1.0, provider-default
\texttt{top\_k}), ensuring consistency across rule extraction tasks. 

\subsubsection{Multi-agent Debate (Architecture B)}
In Architecture~B (\emph{Script Execution Mode}), compliance is determined
through executable rule scripts rather than direct packet-level LLM reasoning.
Here, the \emph{Script Generator Agent} translates extracted rules into a
Python function, which is then reviewed by a \emph{Critic Agent}.
If flaws are found, an iterative debate cycle is launched: the Script Generator
produces a revised script $S^{(t+1)}$ conditioned on the Critic’s feedback of
$S^{(t)}$.  Unlike Architecture~A, where debate is capped at five rounds per
section, the script–critic cycle is permitted up to 20 iterations to ensure
a fully executable and correct script. Runtime errors encountered during
execution are also looped back into this cycle until either a valid script is
produced or the retry cap is reached. Prompt samples for both Architectures A and B have been discussed in \mbox{\textbf{Appendix D}}. The effect of Multi-Agent debate for reasoning has been discussed in \textbf{Appendix E} with important examples for both Architectures.

\subsubsection{LLM providers and access layer}
\label{subsec:llm_providers}

All large language models were accessed exclusively through production-ready APIs. Identical inference parameters were enforced across all experiments with no prompt tuning, ensuring strict comparability across providers. 

A total of sixteen state-of-the-art LLMs were evaluated. Table~\ref{tab:llm_access} provides the complete mapping of model, provider, and API/access layer.

\begin{table}[h]
\centering
\caption{LLM providers, models, and access layers used in experiments}
\label{tab:llm_access}
\begin{tabular}{|l|l|l|}
\hline
\textbf{Provider} & \textbf{Model(s)} & \textbf{API / Access Layer} \\
\hline
\multirow{5}{*}{\textsc{OpenAI}} 
  & GPT-4o & \multirow{5}{*}{OpenAI API \cite{openaiapi}} \\
  & o1 & \\
  & GPT-4-Turbo & \\
  & GPT-4o-mini & \\
  & GPT-3.5-Turbo & \\
\hline
\multirow{3}{*}{\textsc{Meta}} 
  & Llama 3.3-70B & \multirow{3}{*}{Groq API \cite{groqapi}} \\
  & Llama 3.2-11B & \\
  & Llama 3.1-8B & \\
\hline
\multirow{2}{*}{\textsc{Anthropic}} 
  & Claude 3.5 Sonnet & \multirow{2}{*}{Claude API \cite{claudeapi}} \\
  & Claude 4.5 Sonnet & \\
\hline
\multirow{2}{*}{\textsc{Google}} 
  & Gemini 2.0 Flash & \multirow{2}{*}{Gemini API \cite{geminiapi}} \\
  & Gemini 2.5 Flash & \\
\hline
\textsc{Alibaba Cloud} & Qwen-3-32B & Groq API \cite{groqapi} \\
\hline
\textsc{Mistral AI} & Mistral 7B & HuggingFaceHub \cite{huggingfacehub} \\
\hline
\multirow{2}{*}{\textsc{Nvidia--Google}} 
  & Gemma 3-27B & HuggingFaceHub \\
  & Gemma 2-9B & HuggingFaceHub \\
\hline
\end{tabular}
\end{table}

This per-model specification ensures clarity regarding both provider and execution environment, while covering a diverse set of architectures ranging from dense transformers (GPT, Claude, Gemini) to mixture-of-experts (Mixtral) and lightweight transformer variants (Gemma).

\subsection{Ablation Testing Settings}
\label{subsec:ablation_settings}

To evaluate the contribution of the debate mechanism to compliance accuracy, we conducted ablation experiments using only Architecture~A (Explainability Mode) but with the \emph{multi-agent debate phase disabled}. In this setting, each compliance agent independently assessed its assigned section of the RFC, applied the general IPv6 rules to the input packet, and produced a binary \texttt{Yes}/\texttt{No} verdict accompanied by a short rationale. However, unlike the full framework, no subsequent debate or cross-agent reconciliation was permitted: the initial verdict of each agent was treated as final.


\section{Results}
\label{sec:results}
In this section, we present a comprehensive evaluation of our compliance classification framework across multiple architectural modes and experimental settings. The goal is to assess not only the raw performance of individual LLMs but also the contributions of multi-agent debate and rule-execution pipelines to overall system accuracy, robustness, and efficiency. Results are reported using standard metrics of accuracy, F1-score, and latency, supplemented by detailed analysis of model reasoning quality and system-level behavior.

We first conduct an ablation study to isolate the role of the Multi-Agent Debate mechanism by evaluating Architecture~A in \emph{Explainability Mode} with the debate component disabled. This baseline reveals the raw ability of different models to classify compliance based solely on their initial reasoning.

Next, we reintroduce the Multi-Agent Debate mechanism in Architecture~A, measuring how collaborative reasoning improves decision quality. In addition to accuracy, F1, and latency, we also report the number of debate rounds triggered per model. This provides insight into how frequently models disagreed, and how debate dynamics resolved ambiguities. Special attention is given to chain-of-thought (CoT) models \cite{wei2022chain}, which benefit disproportionately from explainability-driven modes.

We then evaluate Architecture~B, where compliance is determined through executable rule scripts rather than direct LLM classification. This mode emphasizes reproducibility and correctness guarantees: we report not only performance metrics but also script retry counts and error recovery rates, which highlight the strengths and limitations of the script–critic loop.

Finally, we present a Comparative Analysis, where we synthesize results across both architectures and benchmark our framework against classical machine learning and deep learning baselines. This comparison situates our system in the broader compliance verification landscape and underscores the advantages of agentic, explainability-driven approaches.

\begin{figure*}[t]
    \centering
    \begin{subfigure}[t]{0.495\textwidth}
        \centering
        \includegraphics[width=\linewidth]{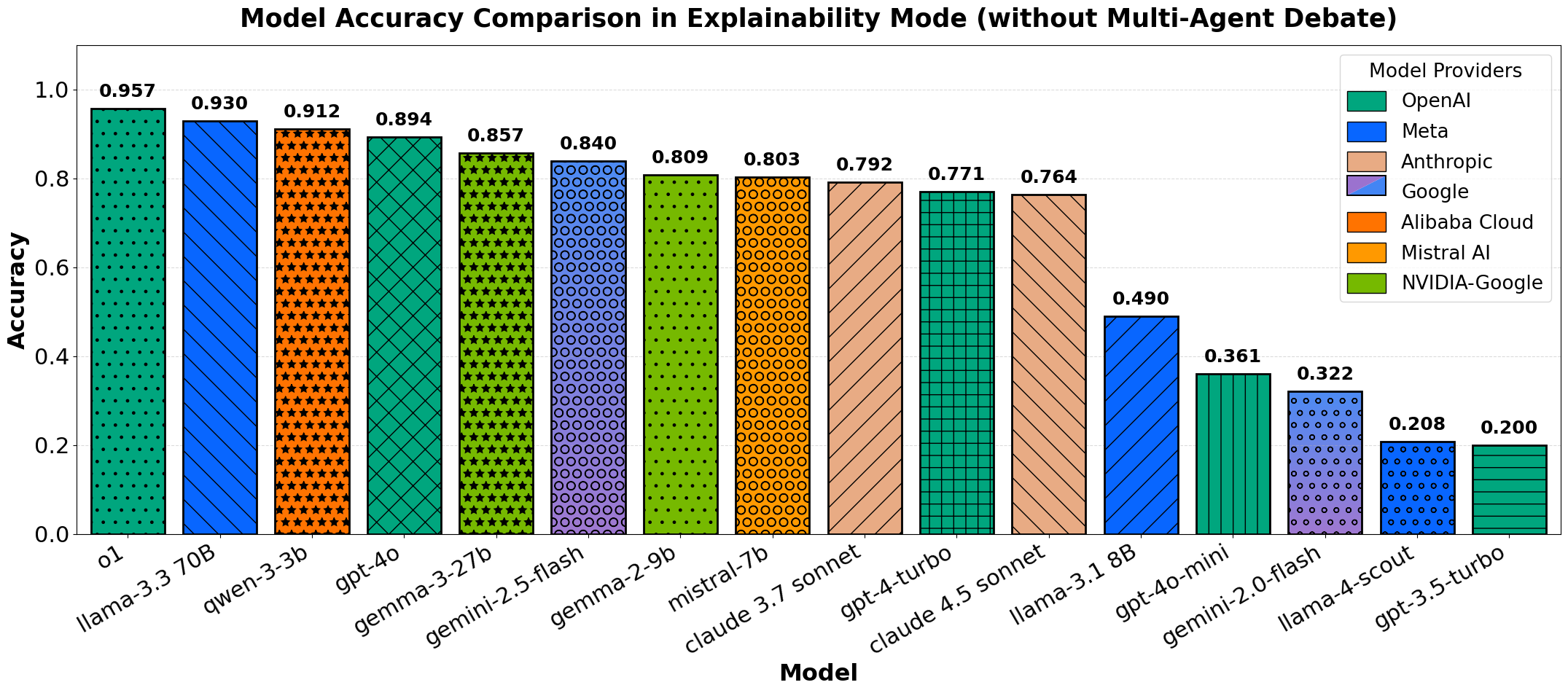}
        \caption{Without Multi-Agent Debate}
    \end{subfigure}
    \hfill
    \begin{subfigure}[t]{0.495\textwidth}
        \centering
        \includegraphics[width=\linewidth]{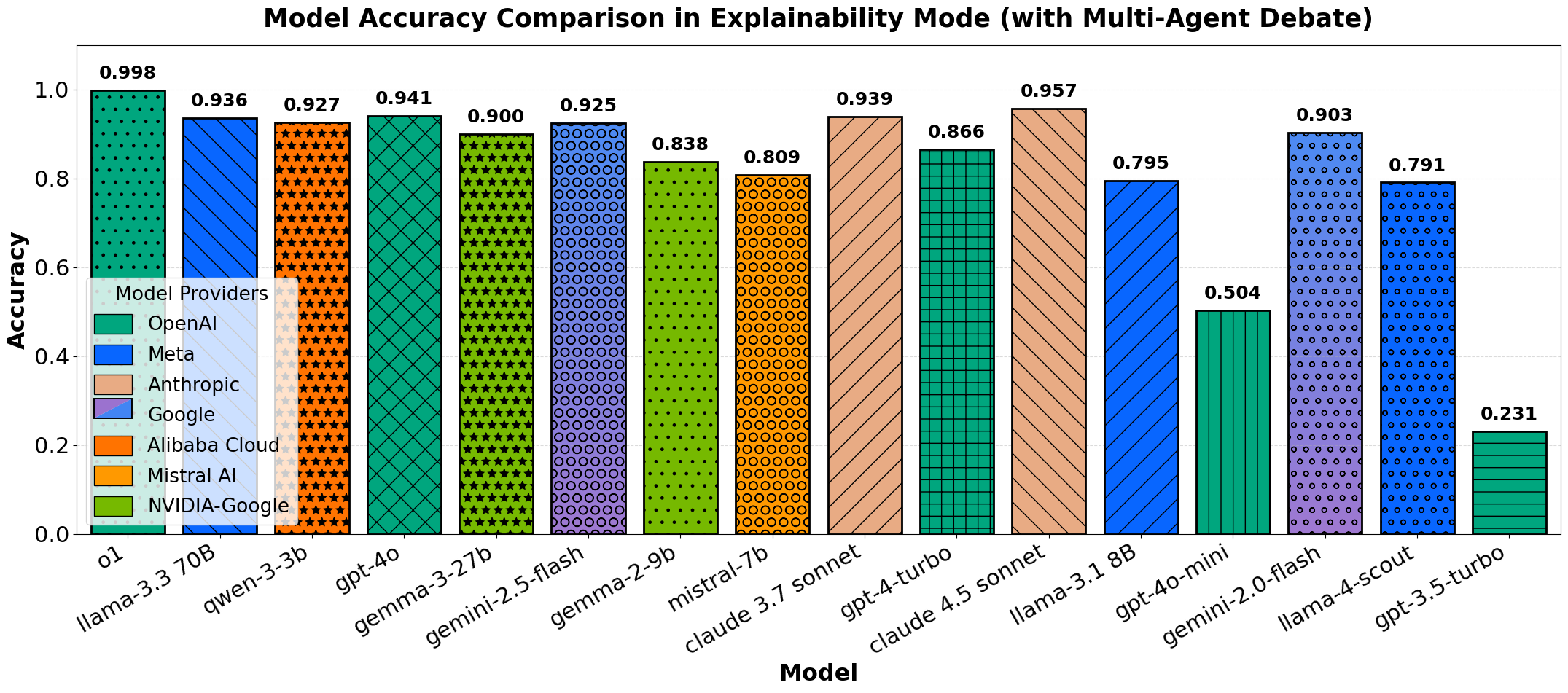}
        \caption{With Multi-Agent Debate}
    \end{subfigure}
    \caption{Model accuracy comparison under ablation vs. debate-enabled settings for Architecture~A. Debate consistently improves accuracy across models, particularly for weaker baselines.}
    \label{fig:acc_compare}
\end{figure*}

\begin{figure*}[t]
    \centering
    \begin{subfigure}[t]{0.495\textwidth}
        \centering
        \includegraphics[width=\linewidth]{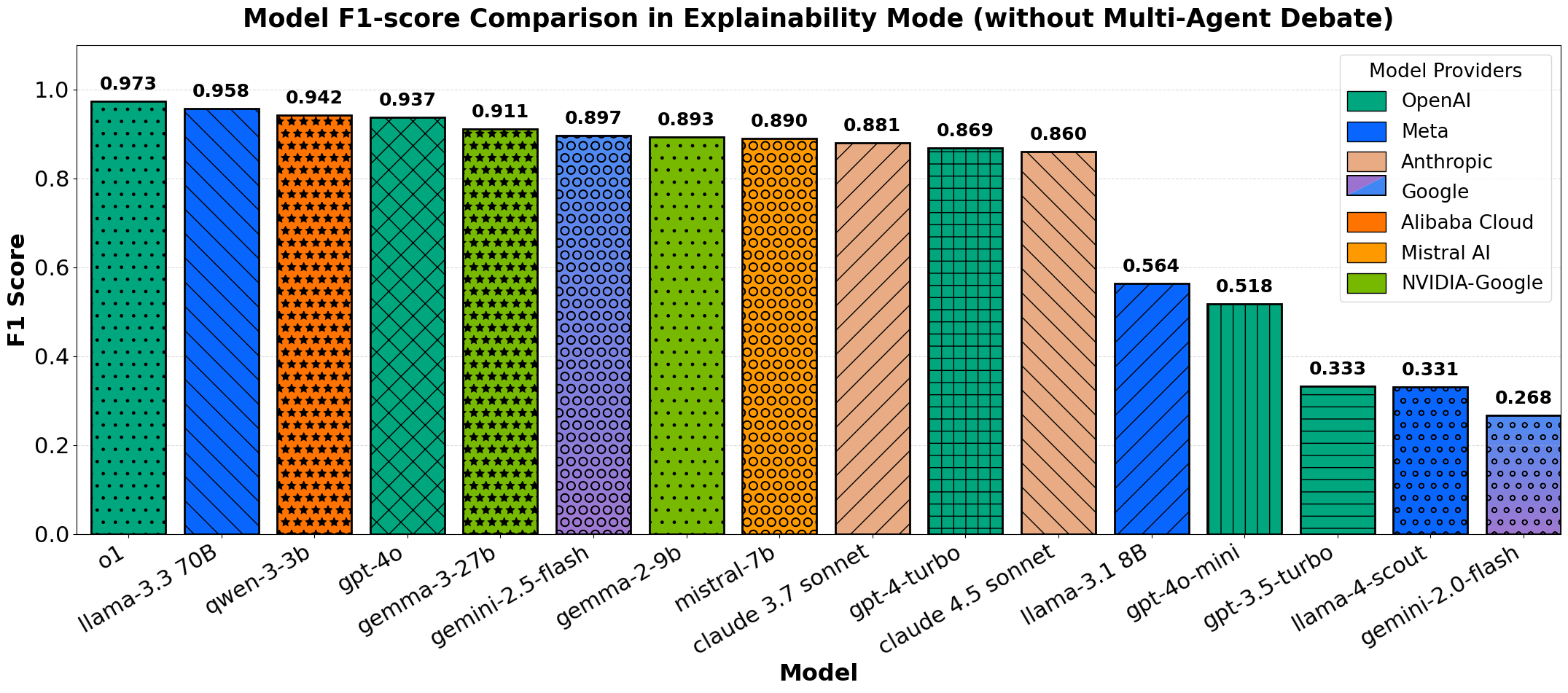}
        \caption{Without Multi-Agent Debate}
    \end{subfigure}
    \hfill
    \begin{subfigure}[t]{0.495\textwidth}
        \centering
        \includegraphics[width=\linewidth]{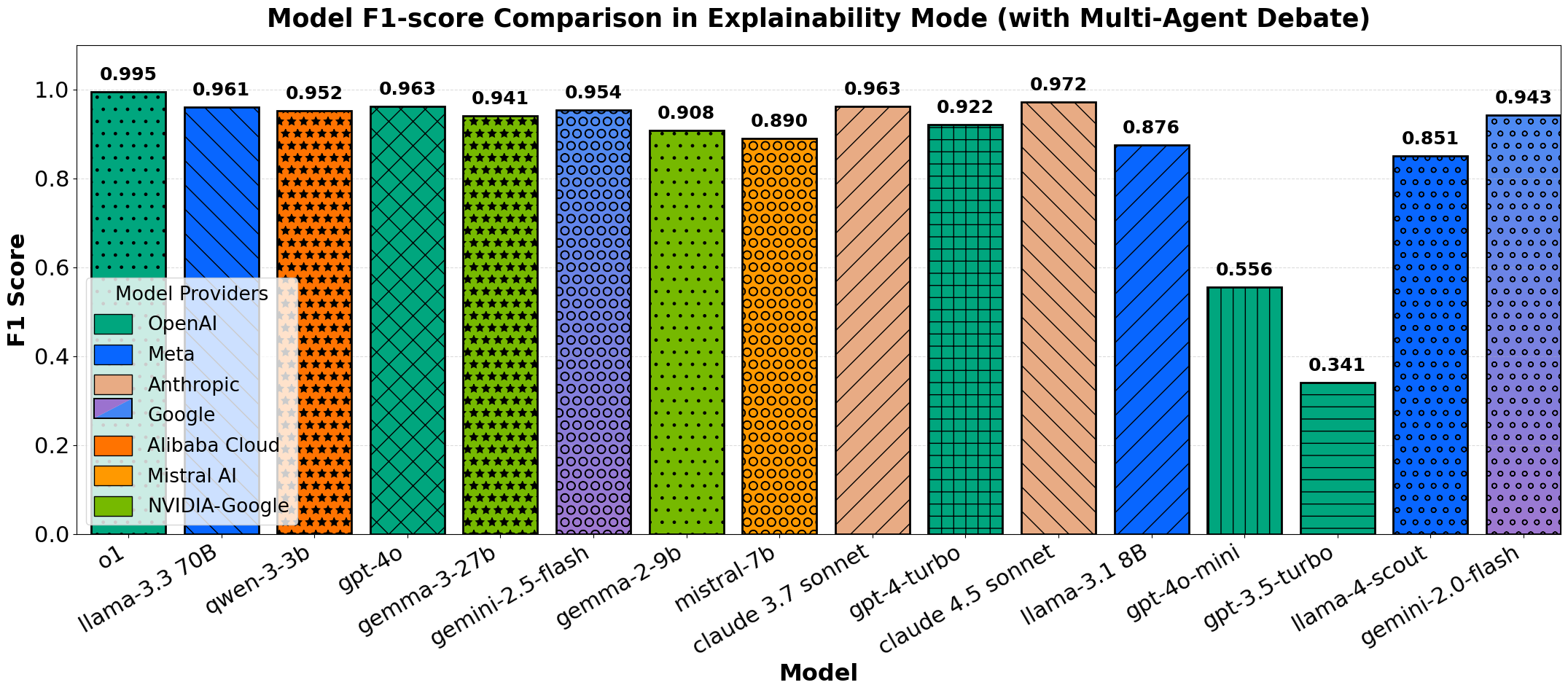}
        \caption{With Multi-Agent Debate}
    \end{subfigure}
    \caption{F1-score comparison under ablation vs. debate-enabled settings for Architecture A. Debate reduces false positives and stabilizes precision–recall balance.}
    \label{fig:f1_compare}
\end{figure*}

\begin{figure*}[t]
    \centering
    \begin{subfigure}[t]{0.495\textwidth}
        \centering
        \includegraphics[width=\linewidth]{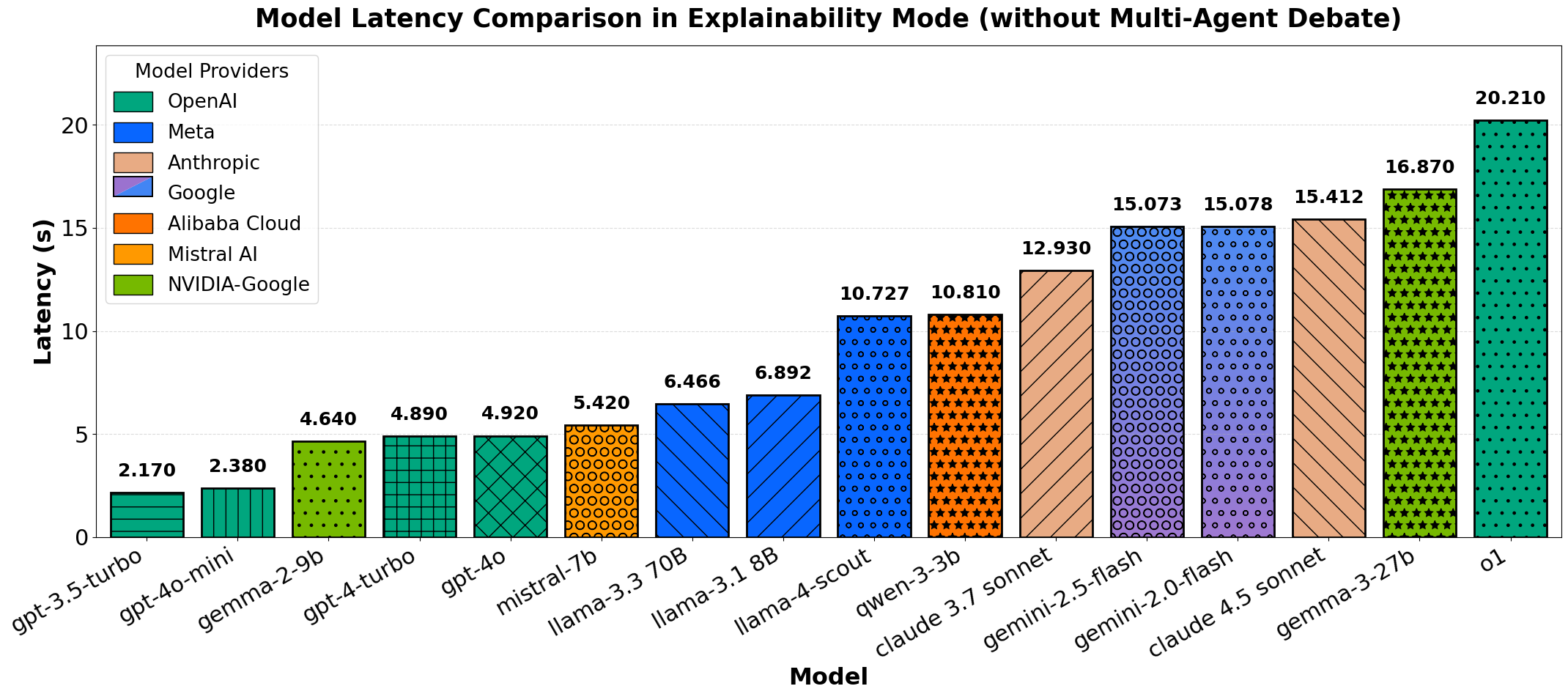}
        \caption{Without Multi-Agent Debate}
    \end{subfigure}
    \hfill
    \begin{subfigure}[t]{0.495\textwidth}
        \centering
        \includegraphics[width=\linewidth]{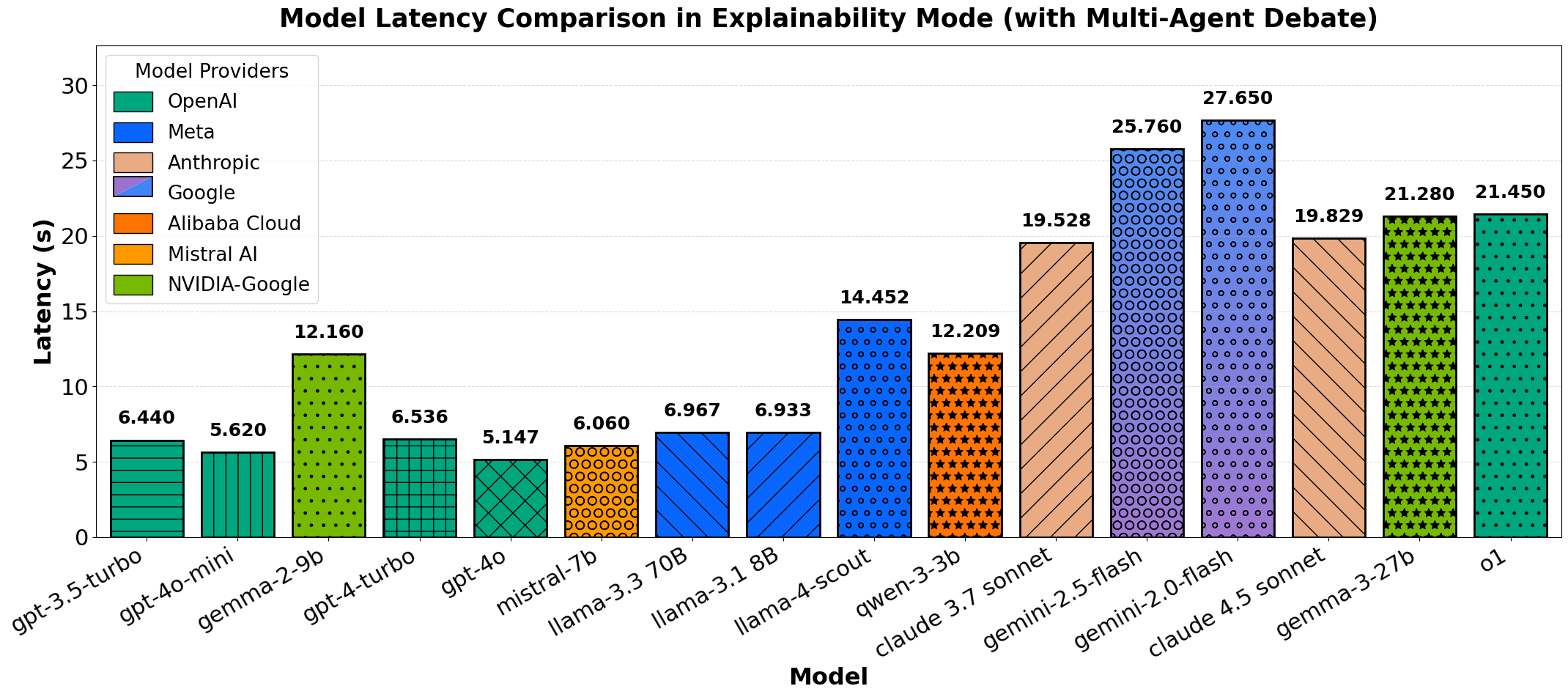}
        \caption{With Multi-Agent Debate}
    \end{subfigure}
    \caption{Decision latency comparison under ablation vs. debate-enabled settings for Architecture A. As expected, enabling debate increases runtime, but accuracy gains justify the overhead.}
    \label{fig:latency_compare}
\end{figure*}

\begin{figure}[t]
    \centering
    \includegraphics[width=0.99\columnwidth]{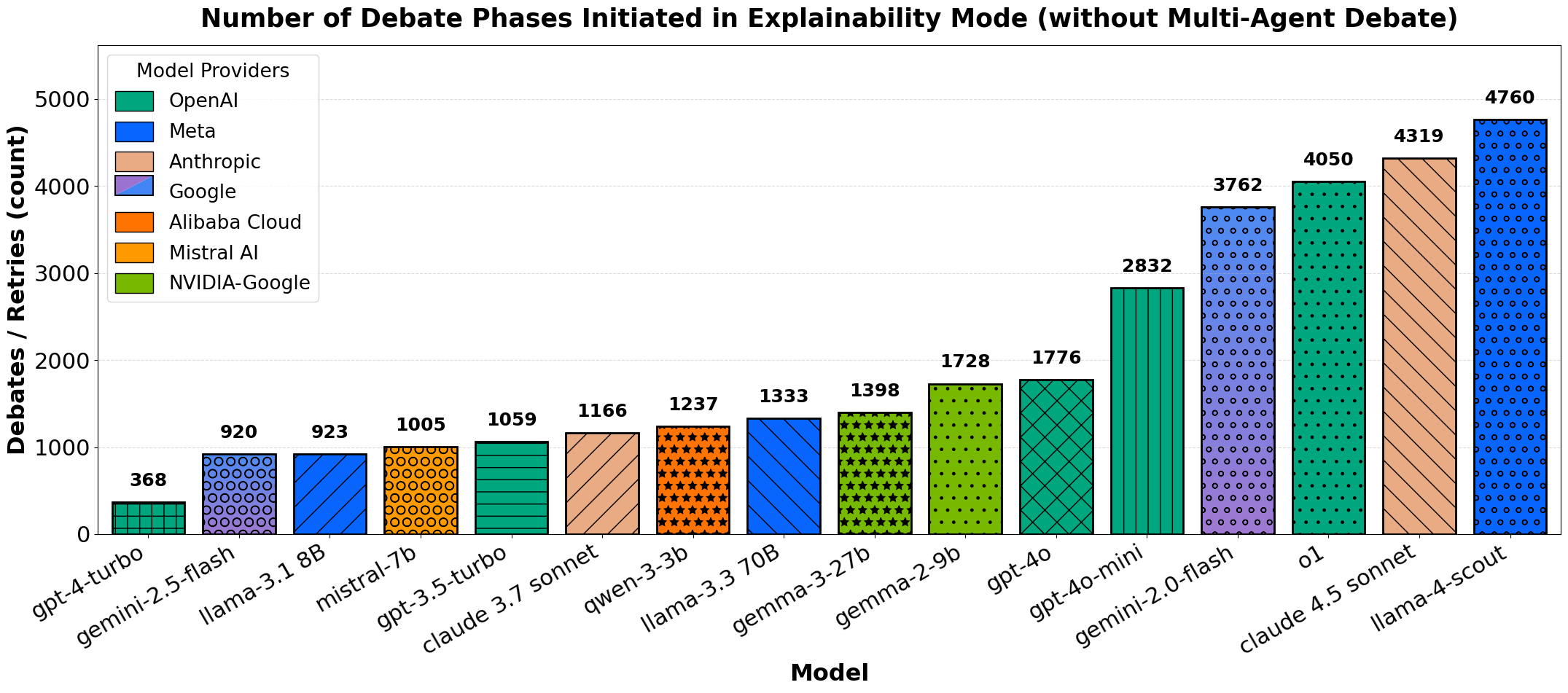}
    \caption{Number of debate phases initiated per model in Architecture~A. Stronger CoT models required fewer debates, while weaker baselines triggered debates more frequently.}
    \label{fig:debate_rounds}
\end{figure}

\subsection{Ablation Testing Results}
\label{sec:ablation_results}

To quantify the impact of the Multi-Agent Debate mechanism, we first evaluate Architecture~A in \emph{Explainability Mode} without the debate component (Ablation Testing). We then reintroduce the debate mechanism and compare results directly. Figures~\ref{fig:acc_compare}, \ref{fig:f1_compare}, and \ref{fig:latency_compare} present side-by-side comparisons of classification accuracy, F1-score, and latency, respectively. In addition, Figure~\ref{fig:debate_rounds} reports the frequency with which debate phases were triggered, illustrating the role of collaborative reasoning in improving verdict quality.

Models with explicit chain-of-thought (CoT) capabilities, such as \texttt{o1} and \texttt{Llama-3.3-70B}, perform strongly even in the ablation setting because their step-by-step reasoning supports more reliable single-pass compliance judgments. Their higher latency is primarily driven by verbose reasoning traces and detailed rationale generation. In contrast, without the multi-agent debate mechanism, models with weaker initial judgment or less stable reasoning cannot revise erroneous decisions, leading non-CoT and lighter-weight models to exhibit substantially lower accuracy and F1 and greater vulnerability to misclassification in the absence of a corrective process.


A closer examination of the models’ explanations illustrates why CoT based models such as o1 outperforms the non-CoT approach. The CoT-based models consistently produces step-by-step rationales for its compliance verdicts. 

For example, consider a packet where the \texttt{Protocol} field is set to ICMP, yet the record clearly contains TCP-related columns, and the Flow Label value is set to \texttt{0xb7cb4a} (decimal 12,035,274). The chain-of-thought (CoT) based model such as o1, produces the following detailed explanation:

\begin{formal}
\noindent \textbf{Verdict: Non-Compliant} \\[4pt]
\textbf{Rationale:} \\
Step 1: The \texttt{Protocol} field is set to ICMP. According to RFC~8200, the \texttt{Protocol} field must correspond to the actual transport-layer header present in the packet. \\ 
Step 2: The record contains TCP-related fields, which directly contradicts the ICMP setting, indicating a protocol mismatch. \\ 
Step 3: The Flow Label is set to \texttt{0xb7cb4a} ($12{,}035{,}274$), which exceeds the 20-bit maximum defined in RFC~8200 (Section~6). Valid Flow Label values must be less than $2^{20} = 1{,}048{,}576$. \\ 
Step 4: Since both conditions violate IPv6 compliance rules, the packet is classified as \emph{Non-Compliant}.
\end{formal}

In contrast, a non-CoT baseline model such as GPT-3.5-Turbo, might return a shorter and less precise justification such as:

\begin{formal}
\noindent \textbf{Verdict: Non-Compliant} \\[4pt]
\textbf{Rationale:} Flow Label exceeds the 20-bit maximum. 
\end{formal}

Here, the non CoT based LLM flagged the Flow Label violation but entirely missed the protocol mismatch caused by the contradictory ICMP/TCP overlap, resulting in an incomplete analysis, although the compliance verdict was correct in this case.

In some instances, the baseline even misclassified borderline compliant cases due to its lack of deeper analysis. By explicitly referring to the relevant criteria in its reasoning, \texttt{o1} not only offers greater transparency but also makes more accurate decisions. This example highlights how chain-of-thought reasoning improves both interpretability and performance in the absence of a debating mechanism.

\subsection{Architecture A: Explainability Mode Results}
\label{sec:explainability_results}

When the Multi-Agent Debate mechanism is enabled, Architecture~A exhibits a marked improvement in both accuracy and F1-score across nearly all evaluated models. Figures~\ref{fig:acc_compare}, \ref{fig:f1_compare}, and \ref{fig:latency_compare} compare performance before and after the introduction of debate, while Figure~\ref{fig:debate_rounds} shows the number of debate phases triggered, offering insight into how disagreement resolution influenced final verdicts.

Overall, Multi-Agent Debate functions as a corrective layer that most strongly benefits models prone to initial misjudgments or unstable reasoning. In particular, weaker baselines such as \texttt{Gemini-2.0-Flash}, \texttt{Llama-4-Scout}, and \texttt{Llama-3.1-8B} become substantially more competitive once debate is enabled, consistent with their high debate activity (i.e., frequent disagreement resolution over the evaluation set). Stronger models such as \texttt{o1}, \texttt{GPT-4o}, \texttt{Llama-3.3-70B}, and \texttt{Qwen-3-3B} exhibit smaller deltas, suggesting their single-pass reasoning is already near-optimal and debate mainly resolves a residual set of ambiguous cases.

At the same time, debate is not theoretically guaranteed to improve every model under every evaluation regime. Its impact depends on the base model’s calibration, the relative strength of debating agents, the debate trigger frequency, and whether additional rounds introduce noise or over-correction on borderline cases. When a model is already near ceiling performance, a small number of debate-induced flips can shift the precision–recall balance and slightly reduce F1, especially in limited-scale evaluations; with larger evaluation sets, these effects typically stabilize and the net corrective benefit becomes more consistent.

Enabling debate increases latency, as expected, but the overhead is generally acceptable given the robustness and reliability gains. High-capacity models tend to incur moderate additional runtime, whereas some hosted APIs exhibit higher absolute latency. Overall, these results motivate Architecture~A when explainability and deliberative reliability are desired, while acknowledging that debate benefits are conditional rather than universal.

\begin{figure*}[t]
    \centering
    \begin{subfigure}[t]{0.495\textwidth}
        \centering
        \includegraphics[width=\linewidth]{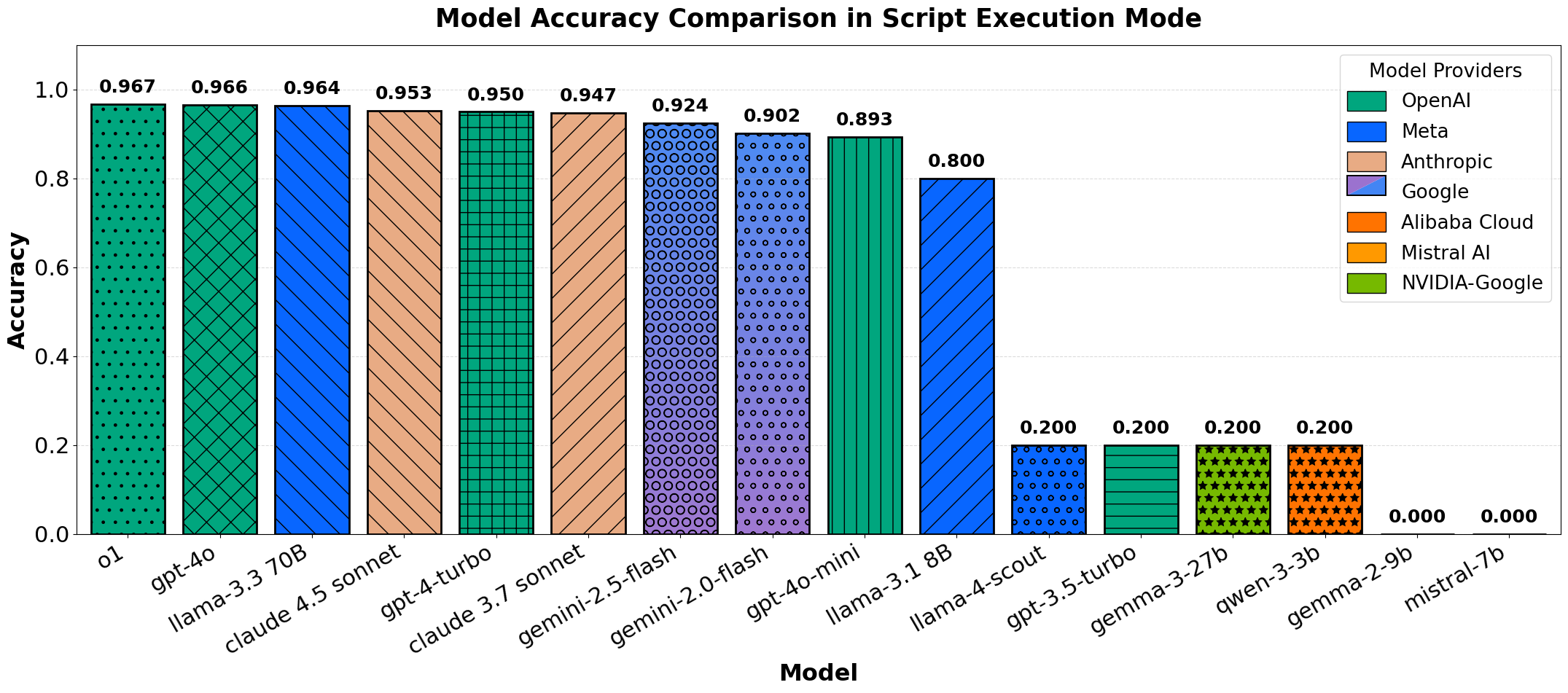}
        \caption{Accuracy}
        \label{fig:acc_script}
    \end{subfigure}
    \hfill
    \begin{subfigure}[t]{0.495\textwidth}
        \centering
        \includegraphics[width=\linewidth]{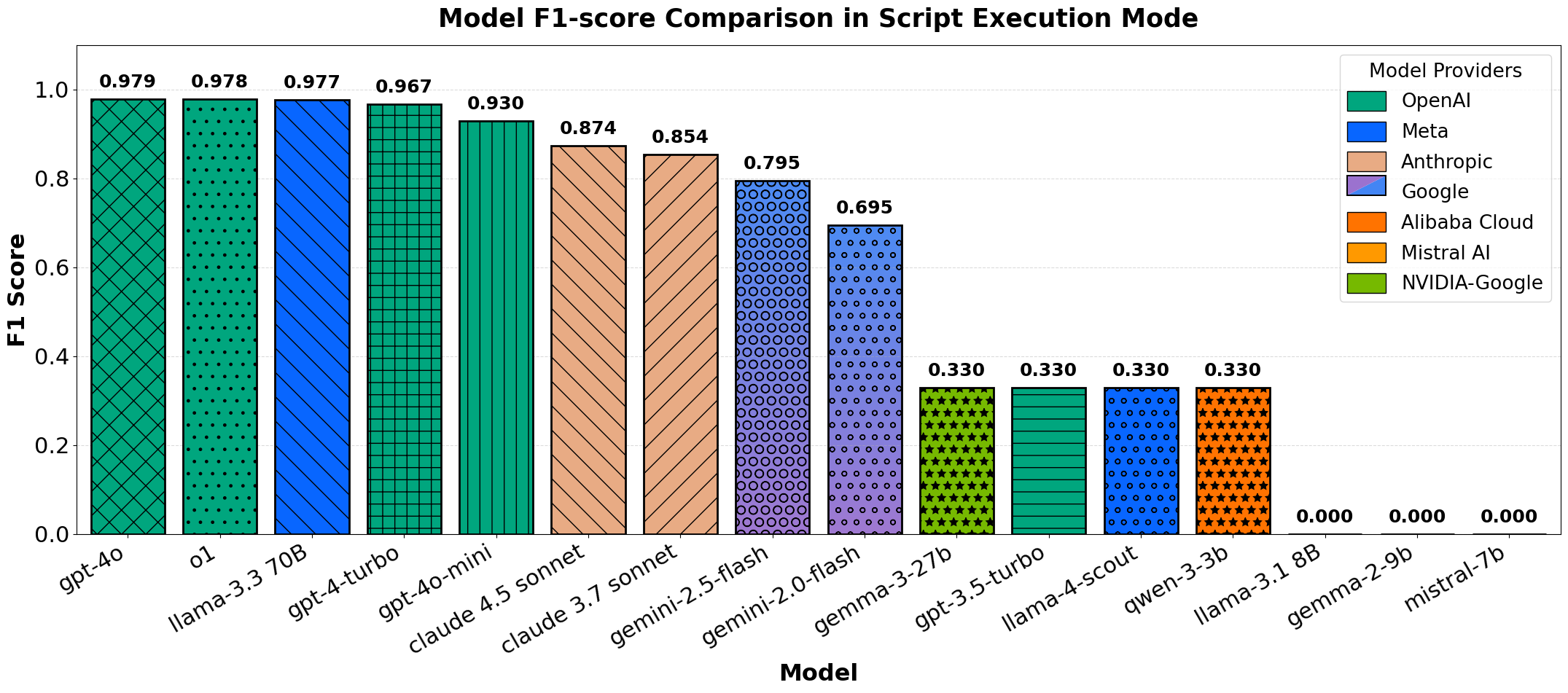}
        \caption{F1-score}
        \label{fig:f1_script}
    \end{subfigure}
    \hfill
    \begin{subfigure}[t]{0.495\textwidth}
        \centering
        \includegraphics[width=\linewidth]{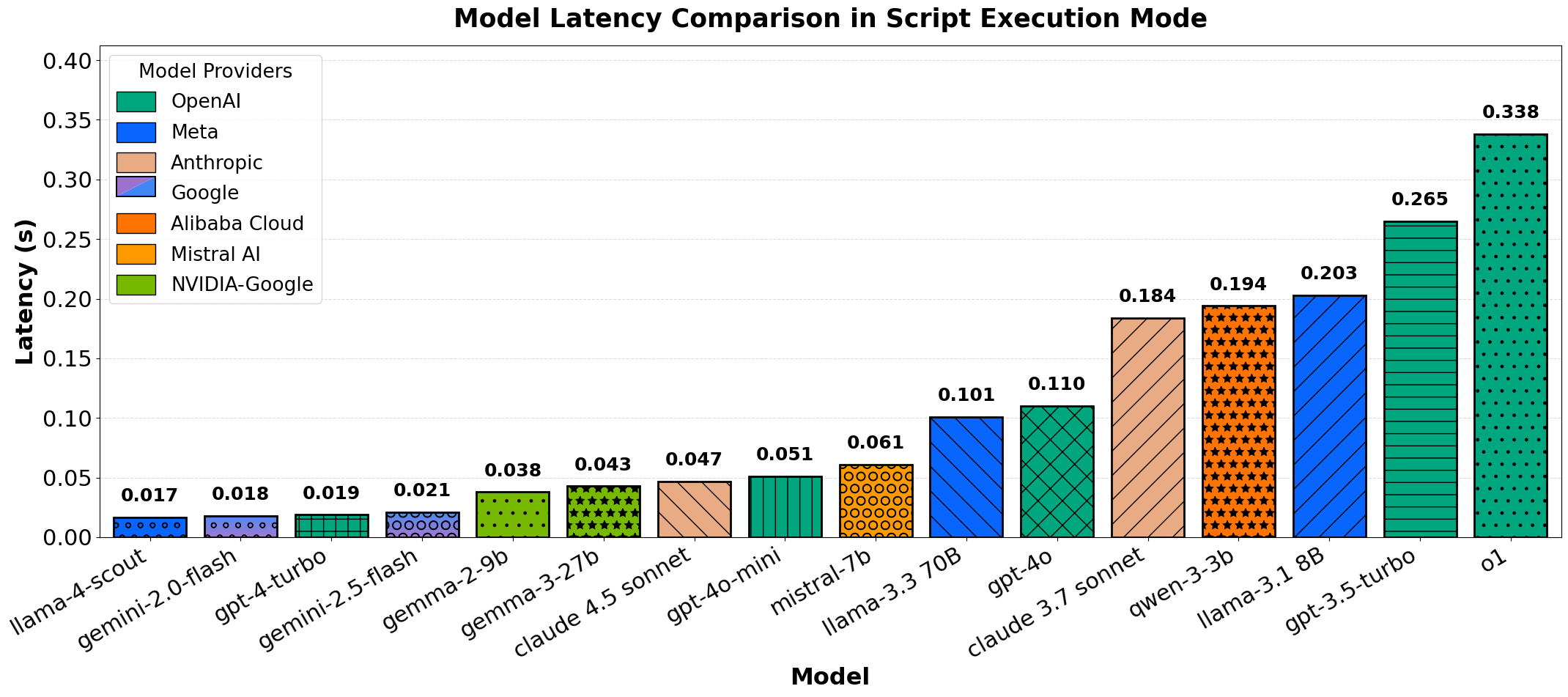}
        \caption{Latency}
        \label{fig:latency_script}
    \end{subfigure}
    \hfill
    \begin{subfigure}[t]{0.495\textwidth}
        \centering
        \includegraphics[width=\linewidth]{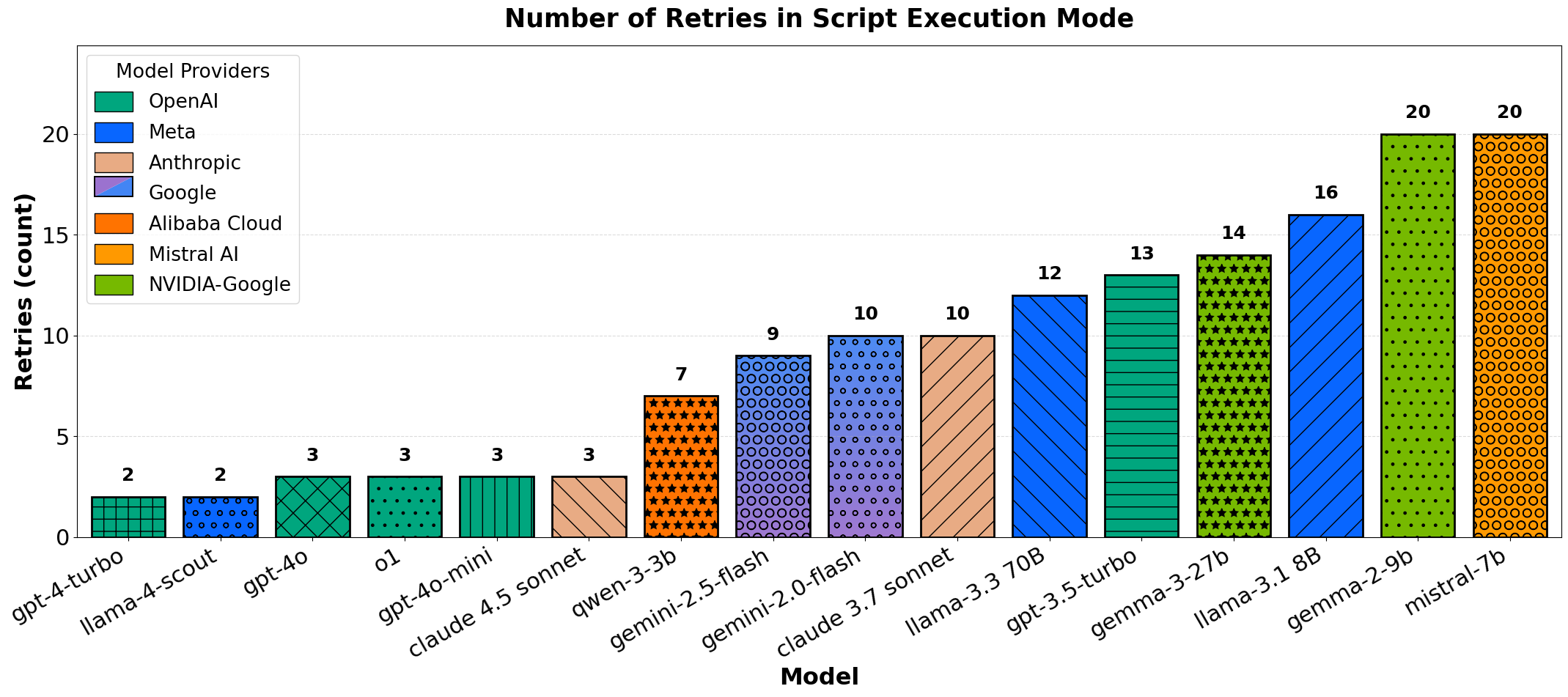}
        \caption{Retries}
        \label{fig:retries_script}
    \end{subfigure}
    \caption{Performance of models in Script Execution Mode (Architecture~B). Subfigures show (a) accuracy, (b) F1-score, (c) latency, and (d) number of retries.}
    \label{fig:script_exec_results}
\end{figure*}

\subsection{Architecture B: Script Execution Results}
\label{sec:script_results}

Architecture~B adopts a fundamentally different approach by shifting from natural language verdict generation to executable compliance scripts. This design enforces stricter consistency, as the models must output structured code that is executed against packet attributes. Figures~\ref{fig:acc_script}, \ref{fig:f1_script}, and \ref{fig:latency_script} report accuracy, F1-score, and latency for all models, while Figure~\ref{fig:retries_script} shows the number of retries required when script generation initially failed.

Overall, the results show that Architecture~B is highly reliable for strong foundation models, which can consistently translate RFC clauses into executable compliance logic under the script-execution verifier. Models such as \texttt{o1}, \texttt{GPT-4o}, \texttt{GPT-4-Turbo}, \texttt{Llama-3.3-70B}, and the stronger \texttt{Claude} variants perform robustly, while several weaker models collapse; such as predicting a single class in an imbalanced setting or fail to produce usable scripts within the retry budget (e.g., \texttt{GPT-3.5-Turbo}, \texttt{Llama-4-Scout}, \texttt{Qwen-3-3B}, \texttt{Gemma-2-9B}, \texttt{Gemma-3-27B}, \texttt{Mistral-7B}), highlighting Architecture~B’s sensitivity to structured code generation quality. These failures highlight the sensitivity of Architecture~B to structured code generation: once a model repeatedly emits syntactically invalid or semantically incorrect scripts, the execution framework provides no opportunity for recovery beyond the retry budget.

Latency results further emphasize Architecture~B’s advantage: because compliance is enforced by executing concise scripts instead of producing long explanations, decision times are typically sub-second across models, making this mode suitable for high-throughput settings. Retry statistics align with these trends, high-performing models generally require few retries, whereas weaker models frequently exhaust the retry budget or repeatedly generate executable but semantically incorrect scripts, demonstrating that the execution framework sharply separates models that can reliably produce correct rule logic from those that cannot.

\begin{table*}[htbp]
\centering
\caption{Performance Comparison of Different Compliance Checker Systems}
\label{tab:system_comparison}
\begin{tabular}{lccccccc}
\toprule
\textbf{System} & \textbf{Accuracy} & \textbf{Precision} & \textbf{Recall} & \textbf{F1-score} & \textbf{Latency} & \textbf{Explainability} & \textbf{Err. Recovery} \\
\midrule
Logistic Regression$^{*}$                           & 0.93  & 0.9452 & 0.69  & 0.7977 & 0.00035 & No  & No \\
Random Forest$^{*}$                                 & 0.938 & 0.8557 & 0.83  & 0.8426 & 0.00086 & No  & No \\
Extra Trees$^{*}$                                   & 0.928 & 0.8333 & 0.80  & 0.8163 & 0.00043 & No  & No \\
XGBoost$^{*}$                                       & 0.56  & 0.2987 & 0.89  & 0.4472 & 0.00016 & No  & No \\
LightGBM$^{*}$                                      & 0.50  & 0.2656 & 0.85  & 0.4048 & 0.00020 & No  & No \\
Support Vector Machine$^{*}$                        & 0.914 & 0.9831 & 0.58  & 0.7296 & 0.00029 & No  & No \\
Gradient Boosting$^{*}$                             & 0.85  & 0.6033 & 0.73  & 0.6606 & 0.00145 & No  & No \\
K-Nearest Neighbors$^{*}$                           & 0.888 & 0.9583 & 0.46  & 0.6216 & 0.00001 & No  & No \\
Naive Bayes$^{*}$                                   & 0.882 & 1.0000 & 0.41  & 0.5816 & 0.00001 & No  & No \\
MLP Neural Net$^{*}$                                & 0.93  & 0.8095 & 0.85  & 0.8293 & 0.00233 & No  & No \\
BERT + LSTM$^{*}$                                   & 0.84  & 1.0000 & 0.20  & 0.3333 & 0.3653  & No  & No \\
Isolation Forest$^{*}$                              & 0.774 & 0.4270 & 0.38  & 0.4021 & 0.00070 & No  & No \\
\textbf{AICCE Arch A (w/o MAD)}               & \textbf{0.957} & 0.964 & 0.982 & 0.973 & 20.210  & \textbf{Yes} & No \\
\textbf{AICCE Arch A (with MAD)}              & \textbf{0.998} & 1.000 & 0.990 & 0.995 & 21.450  & \textbf{Yes} & \textbf{Yes} \\
\textbf{AICCE Arch B}                         & \textbf{0.967} & 0.997 & 0.961 & 0.979 & 0.098   & No  & \textbf{Yes} \\
\bottomrule
\end{tabular}
\vspace{2pt}
\\
{\footnotesize $^{*}$ Trained on 1000 samples and tested on 500 samples.}
\end{table*}

In summary, the updated results confirm that Architecture~B offers an excellent accuracy–latency trade-off when applied to strong foundation models. Unlike Explainability Mode, where multi-agent debate is necessary to iteratively repair reasoning errors, Script Execution Mode enforces correctness at the code level, substantially reducing both variance and runtime for capable models. At the same time, it introduces a sharper divide between models that can reliably produce precise, executable compliance rules and those that cannot, with weaker models collapsing to trivial or unusable behavior. This duality underscores both the promise and the challenge of script-based compliance verification in high-stakes interoperability settings.

\subsection{Comparative Analysis}
\label{sec:comparative}

Table~\ref{tab:system_comparison} contrasts AICCE against classical and deep baselines using accuracy, precision, recall, F1, and latency. For AICCE, we report only the best result within each mode (Architecture~A without Multi-Agent Debate (MAD), Architecture~A, and Architecture~B) in terms of accuracy; full per-model results are deferred. The classical machine learning and sequential baselines are trained on 1000 labeled samples and evaluated on a held-out test set of 500 samples, whereas all AICCE modes are evaluated in a zero-shot regime over the full 1500 sample corpus (train + test), with no task-specific training or fine-tuning required. The complete results for all AICCE configurations and large language models across all modes appear in \textbf{Appendix~B}.

Classical discriminative models (e.g., Logistic Regression, Random Forest, MLP) achieve respectable accuracy ($\approx$0.88–0.94) but exhibit pronounced precision–recall imbalances that depress F1 (e.g., SVM precision $0.9831$ vs.\ recall $0.58$; Naive Bayes precision $1.00$ vs.\ recall $0.41$). Tree ensembles with stronger recall (Random Forest, Extra Trees) improve F1 into the 0.82–0.85 range, while boosting methods (XGBoost, LightGBM) underperform on this task. The sequence model (BERT+LSTM) attains perfect precision but very low recall ($0.20$), indicating systematic under-detection; the anomaly detector (Isolation Forest) similarly lags in both recall and F1. These baselines offer microsecond-scale latencies due to fixed feature pipelines, but none of these models offer Explainability or Error Recovery. Implementation details for all compared systems are provided in \textbf{Appendix~A}.

By contrast, \textbf{AICCE} delivers near-perfect performance across its operating modes while adding explicit compliance guarantees. In Explainability Mode without MAD (Architecture~A, ablation), the best-performing configuration achieves an accuracy of approximately $0.96$, with F1-score around $0.97$ and recall above $0.98$, already surpassing the sequential baseline in both accuracy and F1. When the Multi-Agent Debate mechanism is enabled (Architecture~A with MAD), the strongest configuration reaches near-perfect classification quality with $\text{accuracy}\approx0.998$, $\text{F1}\approx0.995$, and $\text{recall}\approx0.99$, at the cost of higher latency on the order of tens of seconds due to retrieval, explicit reasoning, and debate. This makes Architecture~A particularly suitable for settings where rich natural-language justifications, human-in-the-loop review, and auditability are paramount (e.g., standards development, forensic post-incident analysis, or contested compliance decisions), since every verdict is accompanied by traceable, clause-grounded reasoning and, when needed, multi-agent deliberation.

In Script Execution Mode (Architecture~B), AICCE trades a small amount of peak accuracy for a substantial reduction in response time. The top configuration in this mode attains $\text{accuracy}\approx0.97$ and $\text{F1}\approx0.98$ while reducing average decision time to well below one second, reflecting the efficiency of executing synthesized rule scripts once clauses are compiled. Architecture~B is therefore preferable in high-throughput, production monitoring scenarios (e.g., continuous protocol conformance screening or real-time gatekeeping), where thousands of messages must be processed per second and a concise pass/fail outcome is sufficient. In summary, Architecture~B offers the best accuracy–latency trade-off for automated pipelines, whereas Architecture~A is advantageous when maximum correctness, detailed explanations, and support for human oversight and evolving standards are more critical than raw throughput.

\section{Conclusion}
\label{sec:conclusion}

\subsection{Contributions and Discussion}
This paper presented AICCE, a generative framework for IPv6 standards compliance that uses retrieval-augmented generation and has two complimentary modes of operation. To reach auditable, clause-faithful judgments, Architecture~A (Explainability Mode) combines clause-grounded, multi-agent reasoning and structured discussion whereas Architecture~B (Script Execution Mode) assembles retrieved clauses into executable rules that provide predictable, low-latency compliance decisions. The AICCE achieves near-perfect accuracy and F1 in Explainability Mode (up to 0.998 accuracy and 0.995 F1) and high accuracy in Script Execution Mode (up to 0.967 accuracy and 0.979 F1) across diverse LLM families by embedding IPv6 RFCs into a vectorized corpus and conditioning model behavior on retrieved passages.
This results in transparent decisions whose evidentiary basis can be traced to specific standards text. Another noteworthy feature of the AICCE design is its zero-shot nature where we rely on standards-grounded retrieval and model-agnostic orchestration instead of task-specific fine-tuning. Together, the results of this study show that principled retrieval, agentic deliberation, and executable-rule synthesis can effectively close the gap between formal conformance testing and traditional tooling while maintaining interpretability; appropriate for certification and auditing.

\subsection{Limitations}
However, there are several limitations that merit discussion. The chunking technique and retrieval of documents are crucial to AICCE's performance; missing or fragmented passages can impair reasoning and rule extraction, particularly for cross-referenced clauses. Occasionally, over-corrections suggest that human adjudication may still be necessary for edge cases, and debate helps to reduce, but not always eliminate the ambiguities inherent in evolving RFCs. Significant model variance can be seen in weaker models, which can introduce latency and cost variability by failing deterministically at script synthesis or requiring numerous debate rounds. Even though the evaluation dataset is based on actual traces and is enhanced with a variety of violations, it is not possible to fully capture operational corner cases like infrequent extension-header interactions, vendor actions, or hostile covert-channel designs. 

\subsection{Future Directions}
Future work will focus on enhancing retrieval fidelity and knowledge curation by incorporating errata, drafts, and operational best practices, as well as learning re-rankers tailored to protocol discourse and measuring clause coverage with uncertainty estimates. We plan to integrate verifier–generator hybrids that combine debate with specialized verifiers, confidence calibration, and selective abstention to support human-in-the-loop review when evidence is weak or conflicting. A neuro-symbolic layer that compiles retrieved clauses into intermediate logical forms could provide stronger guarantees for temporal or cross-packet properties. Runtime orchestration can be made adaptive by routing samples between Architectures~A and B according to estimated difficulty, explainability needs, and latency budgets, and by allocating debate or refinement iterations dynamically. With the help of sandboxed execution and policy guardrails, robustness will be put to the test against adversarial packets, parser-evasion strategies, and retrieval or prompt poisoning. In order to promote reproducible research in standards-constrained, clause-faithful reasoning, we also hope to expand the framework to other protocols and regulated domains, support multilingual standards, create distilled/on-device variants for edge settings, and publish benchmarks and audit artifacts annotated with clauses.

To summarize, AICCE shows that executable rule synthesis combined with zero-shot, retrieval-grounded multi-agent reasoning can provide high-accuracy, auditable IPv6 compliance at realistic latencies. The framework provides the basis for reliable AI-driven conformance auditing across contemporary networking and related regulated ecosystems by centering the decision-making process around standards text.

 
\bibliographystyle{IEEEtran}
\bibliography{references}

\clearpage

\appendices
\section{Experimental Settings for Baseline Systems}
\label{appendix:baselines}

This appendix details the baseline systems included in the comparative study, the rationale for their selection, and the concrete configurations used. All baselines were executed on the same machine equipped with a single NVIDIA T4 GPU. A uniform preprocessing pipeline was applied to all methods: empty columns were removed, categorical fields were factorized into integer identifiers, missing values were imputed using the most frequent category, and features were standardized to zero mean and unit variance. Evaluation used a held-out test split; we report accuracy, precision, recall, and F1-score for the \(\text{label}=0\) class to match the main text. Per-sample latency is the average over the test set following a warm-up pass.

\subsection{Logistic Regression}
As a calibrated linear baseline with strong interpretability, logistic regression models the posterior via the logistic link
\begin{equation}
p(y=1\mid \mathbf{x})=\sigma\!\big(\mathbf{w}^\top \mathbf{x}+b\big),
\label{eq:logistic}
\end{equation}
where \(\mathbf{x}\in\mathbb{R}^d\) is the standardized feature vector, \(\mathbf{w}\in\mathbb{R}^d\) are parameters, \(b\in\mathbb{R}\) is a bias, and \(\sigma(z)=1/(1+e^{-z})\). Training minimizes the \(L_2\)-regularized negative log-likelihood with inverse regularization strength \(C=1.0\) (default), and a maximum of 500 iterations to ensure convergence under standardization.

\subsection{Support Vector Machine (RBF)}
To provide a margin-based classifier robust to high dimensionality, we used an SVM with a radial basis function kernel:
\begin{equation}
K(\mathbf{x}_i,\mathbf{x}_j)=\exp\!\big(-\gamma \lVert \mathbf{x}_i-\mathbf{x}_j\rVert_2^2\big),
\label{eq:svm-kernel}
\end{equation}
where \(\gamma>0\) controls the kernel width and \(d\) denotes the number of input features after preprocessing (factorization, imputation, and standardization). We set the soft-margin parameter to \(C=1.0\) and used \(\gamma=1/d\) (a common default under RBF kernels). Class probabilities were obtained via post-hoc calibration on the training folds.

\subsection{K–Nearest Neighbors}
As a non-parametric reference that exploits local geometry, KNN predicts by majority vote among the \(k\) nearest neighbors under Euclidean distance,
\begin{equation}
\hat{y}(\mathbf{x})=\mathrm{mode}\!\big(\{y_i:\mathbf{x}_i\in \mathcal{N}_k(\mathbf{x})\}\big),
\label{eq:knn}
\end{equation}
where \(\mathcal{N}_k(\mathbf{x})\) denotes the \(k\) closest training points to \(\mathbf{x}\). We used \(k=5\), a conventional setting balancing variance and bias for standardized features, with uniform neighbor weights.

\subsection{Gaussian Naive Bayes}
To include a fast probabilistic baseline with variance sensitivity, Gaussian Naive Bayes assumes conditional independence,
\begin{equation}
p(y\mid \mathbf{x}) \propto p(y)\prod_{j=1}^{d} \mathcal{N}\!\big(x_j;\,\mu_{j\mid y},\sigma_{j\mid y}^2\big),
\label{eq:gnb}
\end{equation}
where \(x_j\) is the \(j\)-th feature and \(\mu_{j\mid y},\sigma^2_{j\mid y}\) are class-conditional means and variances estimated from training data. No additional smoothing beyond the Gaussian assumption was applied.

\subsection{Random Forest}
To capture non-linearities and feature interactions while reducing variance through bagging, Random Forest aggregates \(T\) decision trees trained on bootstrap replicates with feature sub-sampling at splits. The ensemble predicts by majority vote
\begin{equation}
\hat{y}=\mathrm{mode}\big(\{h_t(\mathbf{x})\}_{t=1}^{T}\big),
\label{eq:rf}
\end{equation}
where \(h_t\) is the \(t\)-th tree. We used \(T=100\) trees, unlimited depth, minimum samples per split \(=2\) (default), minimum samples per leaf \(=1\) (default), features per split \(=\sqrt{d}\) (default for classification), and a fixed random seed for reproducibility.

\subsection{Extra Trees (Extremely Randomized Trees)}
To test a more aggressively randomized ensemble that further decorrelates trees, Extra Trees uses random thresholds at each split. We employed \(T=100\) trees, unlimited depth, default minimum split/leaf sizes (2/1), and \(\sqrt{d}\) features per split (default). The voting rule is identical to Eq. \eqref{eq:rf}.

\subsection{Gradient Boosting (Decision Trees)}
For a strong additive model on tabular data, gradient boosting fits shallow trees stage-wise to the negative gradient of a differentiable loss. The predictor after \(M\) stages is
\begin{equation}
F_M(\mathbf{x})=\sum_{m=1}^{M}\nu\, f_m(\mathbf{x}),
\label{eq:gbm-additive}
\end{equation}
where \(f_m\) is a weak learner (shallow regression tree) and \(\nu\in(0,1]\) is the learning rate. We used \(M=100\) stages (default), tree depth \(=3\) (default), learning rate \(\nu=0.1\) (default), and full-data boosting (default subsampling off), which offers a robust accuracy–latency balance for medium-scale tabular problems.

\subsection{XGBoost}
To incorporate second-order optimization and explicit regularization, XGBoost constructs an additive model over boosting rounds. Let \mbox{\(F_m(\mathbf{x})\)} denote the ensemble prediction after \mbox{\(m\)} stages, and let \mbox{\(f_m(\mathbf{x})\)} denote the regression tree added at stage \mbox{\(m\)}. The model is updated as:
\begin{equation}
F_m(\mathbf{x}) = F_{m-1}(\mathbf{x}) + f_m(\mathbf{x}).
\end{equation}
At stage \mbox{\(m\)}, XGBoost minimizes a quadratic approximation of the loss around the current prediction \mbox{\(F_{m-1}(\mathbf{x}_i)\)}:
\begin{equation}
\mathcal{L}^{(m)} \approx \sum_{i=1}^{n} \Big[g_i f_m(\mathbf{x}_i) + \tfrac{1}{2} h_i f_m(\mathbf{x}_i)^2\Big] + \Omega(f_m),
\label{eq:xgb}
\end{equation}
where \mbox{\(g_i\)} and \mbox{\(h_i\)} are the first and second derivatives of the objective evaluated at \mbox{\(F_{m-1}(\mathbf{x}_i)\)}, and \mbox{\(\Omega\)} penalizes tree complexity (e.g., number of leaves and \mbox{\(L_2\)} leaf weights). We used learning rate \mbox{\(=0.3\)} (default), maximum depth \mbox{\(=6\)} (default), subsample \mbox{\(=1.0\)} (default), column-sample by tree \mbox{\(=1.0\)} (default), minimum child weight \mbox{\(=1\)} (default), \mbox{\(L_2\)} regularization \mbox{\(\lambda=1\)} (default), and the binary logistic objective with log-loss monitoring.

\subsection{LightGBM}
To evaluate histogram-based boosting with leaf-wise growth and efficient feature binning, LightGBM was configured with the standard defaults suitable for balanced accuracy–latency trade-offs: number of leaves \(=31\) (default), maximum depth \(=-1\) (unconstrained; default), learning rate \(=0.1\) (default), feature fraction \(=1.0\) (default), bagging fraction \(=1.0\) (default), and default minimum data in leaf. The underlying additive model follows Eq. \eqref{eq:gbm-additive}.

\subsection{Multi-Layer Perceptron (Feedforward Network)}
As a compact neural baseline for non-linear decision boundaries on standardized tabular features, the MLP comprised a single hidden layer with rectified linear units:
\begin{equation}
\mathbf{h}=\phi\!\big(\mathbf{W}^{(1)}\mathbf{x}+\mathbf{b}^{(1)}\big),
\label{eq:mlp-hidden}
\end{equation}
followed by a two-way softmax:
\begin{equation}
\hat{\mathbf{y}}=\mathrm{softmax}\!\big(\mathbf{W}^{(2)}\mathbf{h}+\mathbf{b}^{(2)}\big),
\label{eq:mlp-softmax}
\end{equation}
where \(\phi\) is ReLU and \(\mathbf{W}^{(1)},\mathbf{W}^{(2)},\mathbf{b}^{(1)},\mathbf{b}^{(2)}\) are learned parameters. The hidden layer used 100 units. Training used the Adam optimizer with a constant learning rate of \(10^{-3}\) and \(\ell_{2}\) weight decay of \(10^{-4}\). Mini-batches contained up to 200 samples (or the full dataset if smaller). Optimization ran for at most 300 epochs with early stopping when the training loss failed to improve by at least \(10^{-4}\) over 10 consecutive epochs.

\subsection{BERT + BiLSTM (Sequence Model)}
To probe whether generic language representations over templated packet text can recover compliance signals, each table row was rendered as a sentence of attribute–value pairs (e.g., “\emph{FieldA is v. FieldB is w.}”), tokenized to a maximum of 128 subwords, and passed through a frozen large uncased Transformer encoder to obtain contextual token embeddings. A bidirectional LSTM then summarized these embeddings, and a linear layer produced logits:
\begin{equation}
\mathbf{z}=\mathbf{W}\,[\overrightarrow{\mathbf{h}}_{T};\overleftarrow{\mathbf{h}}_{1}]+\mathbf{b},
\label{eq:bert-lstm}
\end{equation}
where \(\overrightarrow{\mathbf{h}}_{T}\) and \(\overleftarrow{\mathbf{h}}_{1}\) are the final forward and backward LSTM states, \([\cdot;\cdot]\) denotes concatenation, and \(\mathbf{W},\mathbf{b}\) are classifier parameters. The LSTM used hidden size \(=128\), three layers, and dropout \(=0.3\); batch size \(=32\); learning rate \(=2\times 10^{-5}\); and 16 epochs on the T4. Freezing the encoder isolates the contribution of the light sequence head under identical text rendering.

\subsection{Isolation Forest (Unsupervised Anomaly Detector)}
To assess whether violations appear as statistical outliers, Isolation Forest builds \(T\) random partition trees; anomalies are expected to have shorter average path lengths. The normalized anomaly score is
\begin{equation}
s(\mathbf{x})=2^{-\frac{\mathbb{E}[\ell(\mathbf{x})]}{c(n)}},
\label{eq:isoforest}
\end{equation}
where \(\ell(\mathbf{x})\) is the path length to isolate \(\mathbf{x}\), \(n\) is the subsample size per tree, and \(c(n)\) normalizes for tree height. We used \(T=100\) trees (default) with subsample size \(n=256\) per tree (common default), and set the decision threshold to match the empirical violation rate, yielding a fair precision–recall operating point rather than relying on an automatic contamination heuristic.

\subsection{Timing and Calibration Protocol}
Latencies reflect average per-sample inference time on the test set after a warm-up pass to amortize initialization overheads. Decision thresholds for probabilistic classifiers were \(0.5\). Probability calibration followed each family’s standard approach: intrinsic logistic link for logistic regression and boosted trees trained with logistic losses, post-hoc calibration for SVMs with RBF kernels, and direct posterior outputs where natively defined (Naive Bayes). No per-model feature engineering or prompt tuning was performed; all methods consumed the same standardized inputs so that performance differences reflect modeling assumptions rather than data handling.

\section{Complete Results}
\label{appendix:fullresults}

This appendix reports the full set of quantitative results across all evaluated systems, including every AICCE configuration (both architectures and all large language models) and all classical/deep learning baselines. Table~\ref{tab:aicce_full_simple} consolidates accuracy (Acc), precision (Prec), recall (Rec), F1-score (F1), and average per-sample latency (in seconds). Latency reflects end-to-end wall-clock time per sample after a warm-up pass on the same machine; values include model inference and, for AICCE, the overheads of retrieval, agent orchestration, and (when applicable) debate or script execution.

The reported results are not all produced under the same learning paradigm. The classical machine learning and deep learning baselines are evaluated in a standard supervised setting: they are trained on 1,000 labeled packets and tested on a held-out set of 500 packets. In contrast, all AICCE configurations (both Architecture~A and Architecture~B) are evaluated zero-shot (i.e., without task-specific training or fine-tuning) and are executed over the full 1,500-packet corpus, reflecting end-to-end clause-grounded compliance checking performance without a learned decision boundary.

\begin{table*}[!t]
\centering
\renewcommand{\arraystretch}{1.05}
\caption{Comprehensive comparison across AICCE configurations and baselines.}
\label{tab:aicce_full_simple}
\begin{tabular}{p{2.9cm} l r r r r r c c c}
\toprule
Method & Model & Acc & Prec & Rec & F1 & Latency &
\begin{tabular}[c]{@{}c@{}}Debates\\/Retries\end{tabular} &
\makecell{Offers\\Explainability} & \makecell{Error\\Recovery} \\
\midrule

\multirow{11}{*}{Classical ML$^{*}$} &
Random Forest          & 0.938 & 0.8557 & 0.83 & 0.8426 & 0.00086 & N/A & No & No \\
& MLP Neural Net         & 0.93  & 0.8095 & 0.85 & 0.8293 & 0.00233 & N/A & No & No \\
& Logistic Regression    & 0.93  & 0.9452 & 0.69 & 0.7977 & 0.00035 & N/A & No & No \\
& Extra Trees            & 0.928 & 0.8333 & 0.80 & 0.8163 & 0.00043 & N/A & No & No \\
& SVM                    & 0.914 & 0.9831 & 0.58 & 0.7296 & 0.00029 & N/A & No & No \\
& KNN                    & 0.888 & 0.9583 & 0.46 & 0.6216 & \textbf{0.00001} & N/A & No & No \\
& Naive Bayes            & 0.882 & \textbf{1.0000} & 0.41 & 0.5816 & \textbf{0.00001} & N/A & No & No \\
& Gradient Boosting      & 0.85  & 0.6033 & 0.73 & 0.6606 & 0.00145 & N/A & No & No \\
& Isolation Forest       & 0.774 & 0.4270 & 0.38 & 0.4021 & 0.00070 & N/A & No & No \\
& XGBoost                & 0.56  & 0.2987 & 0.89 & 0.4472 & 0.00016 & N/A & No & No \\
& LightGBM               & 0.50  & 0.2656 & 0.85 & 0.4048 & 0.00020 & N/A & No & No \\
\midrule
\multirow{1}{*}{Sequential Model$^{*}$} &
BERT + LSTM & 0.84 & \textbf{1.00} & 0.20 & 0.3333 & 0.3653 & N/A & No & No \\
\midrule
\multirow{16}{*}{\makecell{AICCE (Arch A \\ without M.A.D)}} &
o1                 & 0.957 & 0.964 & 0.982 & 0.973  & 20.21  & \textbf{0} & \textbf{Yes} & No \\
& gpt-4o            & 0.894 & 0.883 & 0.998 & 0.937  & 4.92   & \textbf{0} & \textbf{Yes} & No \\
& gpt-4-turbo       & 0.771 & 0.797 & 0.956 & 0.869  & 4.89   & \textbf{0} & \textbf{Yes} & No \\
& gpt-3.5-turbo     & 0.200 & 0.200 & \textbf{1.000} & 0.333  & 2.17   & \textbf{0} & \textbf{Yes} & No \\
& gpt-4o-mini       & 0.361 & 0.652 & 0.430 & 0.518  & 2.38   & \textbf{0} & \textbf{Yes} & No \\
& llama-3.3 70B     & 0.930 & 0.922 & 0.997 & 0.958  & 6.466  & \textbf{0} & \textbf{Yes} & No \\
& llama-3.1 8B      & 0.490 & 0.891 & 0.412 & 0.564  & 6.892  & \textbf{0} & \textbf{Yes} & No \\
& llama-4-scout     & 0.208 & 0.199 & 0.980 & 0.331  & 10.727 & \textbf{0} & \textbf{Yes} & No \\
& claude 3.7 sonnet & 0.792 & 0.810 & 0.966 & 0.881  & 12.93  & \textbf{0} & \textbf{Yes} & No \\
& claude 4.5 sonnet & 0.764 & 0.820 & 0.911 & 0.860  & 15.412 & \textbf{0} & \textbf{Yes} & No \\
& gemini-2.5-flash  & 0.840 & 0.924 & 0.871 & 0.897  & 15.073 & \textbf{0} & \textbf{Yes} & No \\
& gemini-2.0-flash  & 0.322 & 0.984 & 0.155 & 0.2678 & 15.078 & \textbf{0} & \textbf{Yes} & No \\
& gemma-3-27b       & 0.857 & 0.890 & 0.925 & 0.911  & 16.87  & \textbf{0} & \textbf{Yes} & No \\
& gemma-2-9b        & 0.809 & 0.807 & \textbf{1.000} & 0.893  & 4.64   & \textbf{0} & \textbf{Yes} & No \\
& qwen-3-3b         & 0.912 & 0.993 & 0.896 & 0.942  & 10.81  & \textbf{0} & \textbf{Yes} & No \\
& mistral-7b        & 0.803 & 0.802 & \textbf{1.000} & 0.890  & 5.42   & \textbf{0} & \textbf{Yes} & No \\
\midrule
\multirow{16}{*}{AICCE (Arch A)} &
o1                 & \textbf{0.998} & \textbf{1.000} & 0.990 & \textbf{0.995} & 21.45  & 4050  & \textbf{Yes} & \textbf{Yes} \\
& gpt-4o            & 0.941 & 0.956 & 0.970 & 0.963 & 5.147  & 1776  & \textbf{Yes} & \textbf{Yes} \\
& gpt-4-turbo       & 0.866 & 0.861 & 0.992 & 0.922 & 6.536  & 368   & \textbf{Yes} & \textbf{Yes} \\
& gpt-3.5-turbo     & 0.231 & 0.205 & 0.995 & 0.341 & 6.44   & 1059 & \textbf{Yes} & \textbf{Yes} \\
& gpt-4o-mini       & 0.504 & 0.981 & 0.387 & 0.556 & 5.62   & 2832  & \textbf{Yes} & \textbf{Yes} \\
& llama-3.3 70B     & 0.936 & 0.928 & 0.996 & 0.961 & 6.967  & 1333  & \textbf{Yes} & \textbf{Yes} \\
& llama-3.1 8B      & 0.795 & 0.844 & 0.911 & 0.876 & 6.933  & 923   & \textbf{Yes} & \textbf{Yes} \\
& llama-4-scout     & 0.791 & 0.990 & 0.746 & 0.851 & 14.452 & 4760  & \textbf{Yes} & \textbf{Yes} \\
& claude 3.7 sonnet & 0.939 & 0.931 & 0.997 & 0.963 & 19.528 & 1166  & \textbf{Yes} & \textbf{Yes} \\
& claude 4.5 sonnet & 0.957 & 0.988 & 0.957 & 0.972 & 19.829 & 4319  & \textbf{Yes} & \textbf{Yes} \\
& gemini-2.5-flash  & 0.925 & 0.927 & 0.983 & 0.954 & 25.76  & 920   & \textbf{Yes} & \textbf{Yes} \\
& gemini-2.0-flash  & 0.903 & 0.895 & 0.995 & 0.943 & 27.65  & 3762  & \textbf{Yes} & \textbf{Yes} \\
& gemma-3-27b       & 0.900 & 0.889 & \textbf{1.000} & 0.941 & 21.28  & 1398  & \textbf{Yes} & \textbf{Yes} \\
& gemma-2-9b        & 0.838 & 0.8316 & \textbf{1.000} & 0.9081 & 12.16 & 1728  & \textbf{Yes} & \textbf{Yes} \\
& qwen-3-3b         & 0.927 & 0.990 & 0.917 & 0.952 & 12.209 & 1237  & \textbf{Yes} & \textbf{Yes} \\
& mistral-7b        & 0.809 & 0.825 & 0.966 & 0.890 & 6.06   & 1005  & \textbf{Yes} & \textbf{Yes} \\
\midrule
\multirow{16}{*}{AICCE (Arch B)} &
o1                 & 0.967 & 0.997 & 0.961 & 0.978 & 0.338   & 3  & No & \textbf{Yes} \\
& gpt-4o            & 0.966 & 0.959 & \textbf{1.000} & 0.979 & 0.11    & 3  & No & \textbf{Yes} \\
& gpt-4-turbo       & 0.950 & 0.992 & 0.945 & 0.967 & 0.01894 & 2  & No & \textbf{Yes} \\
& gpt-3.5-turbo     & 0.200 & 0.200 & \textbf{1.000} & 0.330 & 0.265   & 13 & No & \textbf{Yes} \\
& gpt-4o-mini       & 0.893 & 0.984 & 0.880 & 0.930 & 0.051   & 3  & No & \textbf{Yes} \\
& llama-3.3 70B     & 0.964 & 0.956 & \textbf{1.000} & 0.977 & 0.101   & 12 & No & \textbf{Yes} \\
& llama-3.1 8B      & 0.800 & 0.000 & 0.000 & 0.000 & 0.203   & 16 & No & \textbf{Yes} \\
& llama-4-scout     & 0.200 & 0.200 & \textbf{1.000} & 0.330 & 0.0167  & 2  & No & \textbf{Yes} \\
& claude 3.7 sonnet & 0.947 & 0.945 & 0.780 & 0.854 & 0.184   & 10 & No & \textbf{Yes} \\
& claude 4.5 sonnet & 0.953 & 0.937 & 0.820 & 0.874 & 0.047   & 3  & No & \textbf{Yes} \\
& gemini-2.5-flash  & 0.924 & 0.864 & 0.735 & 0.795 & 0.021   & 9  & No & \textbf{Yes} \\
& gemini-2.0-flash  & 0.902 & 0.918 & 0.560 & 0.695 & 0.018   & 10 & No & \textbf{Yes} \\
& gemma-3-27b       & 0.200 & 0.200 & \textbf{1.000} & 0.330 & 0.043   & 14 & No & \textbf{Yes} \\
& gemma-2-9b        & 0.000 & 0.000 & 0.000 & 0.000 & 0.038   & 20 & No & \textbf{Yes} \\
& qwen-3-3b         & 0.200 & 0.200 & \textbf{1.000} & 0.330 & 0.194   & 7  & No & \textbf{Yes} \\
& mistral-7b        & 0.000 & 0.000 & 0.000 & 0.000 & 0.061   & 20 & No & \textbf{Yes} \\
\bottomrule
\end{tabular}

\vspace{2pt}
{\footnotesize $^{*}$Trained on 1000 samples and tested on 500 samples.}
\end{table*}

The column \emph{Debates/Retries} quantifies, for AICCE Architecture~A, the total number of debate phases invoked across the 1500-sample evaluation (a proxy for how frequently agents disagreed and were reconciled), and for Architecture~B, the number of code-generation retries in the script–critic loop (with “20 (max)” indicating the retry cap was reached without producing a valid script). Classical and deep baselines do not perform either process and are therefore marked “N/A”. The columns \emph{Offers Explainability} and \emph{Error Recovery} indicate whether a system provides clause-grounded natural-language rationales (Explainability) and whether it can automatically correct an initial failure via structured debate (Architecture~A) or iterative script repair (Architecture~B).

\section{Compliance Violation Types}
\label{appendix:violations}

This appendix catalogs the violation categories used to construct and evaluate non-compliant IPv6 samples. Each item states the governing constraint at the header or subheader level and describes the condition under which a packet is deemed non-compliant. The taxonomy was selected to span syntactic errors (e.g., illegal bit-widths), semantic inconsistencies (e.g., payload–header mismatches), and fields commonly exploited for evasion or covert signaling.

\subsection{IPv6 Version Violation}
The IPv6 header includes a fixed 4-bit version field that must equal 6. Any deviation (e.g., values of 4 or 5) reflects a fundamental protocol misuse, typically indicative of malformed traffic or intentional evasion.

\subsection{Length Field Violation}
The \emph{Payload Length} field is 16\,bits and encodes the number of octets after the IPv6 base header (i.e., it excludes the 40\,byte header). In ordinary packets it must lie in \([1,\,65{,}535]\). A value of \(0\) is reserved exclusively for jumbograms; in that case the actual payload length (which may exceed \(65{,}535\) bytes) is carried in the Jumbo Payload option of the Hop-by-Hop Options header.

Violations include any mismatch between the \emph{Payload Length} value and the actual payload size; use of \(0\) without a valid Jumbo Payload option (or a nonzero value when that option is present); and packets whose total size (header \(+\) payload) exceeds the link MTU (Maximum Transmission Unit) or the discovered Path~MTU. These conditions indicate corruption, misconfiguration, or deliberate evasion attempts.

\subsection{IPv6 Address Violation}
Source and destination addresses must follow colon–hexadecimal notation with eight 16-bit blocks and permitted compression (\texttt{::}). Invalid formats (e.g., illegal characters, truncated groups, or missing fields) are syntactically incorrect and non-routable.

\subsection{Hop Limit Violation}
The hop limit (analogous to IPv4 TTL) must be in \([1,255]\) and decrements by one at each router. Values of 0, negative encodings, or excessively large numbers indicate malformed packets that cannot be reliably forwarded.

\subsection{TCP/UDP/ICMPv6 Overlap}
The \emph{Protocol} must indicate exactly one upper-layer protocol. Records that simultaneously populate TCP, UDP, or ICMPv6 header fields are structurally invalid. Although uncommon in benign traffic, such overlaps can be crafted for covert communication or parser evasion and are treated as non-compliant.

\subsection{Flow Label Violation}
The \emph{Flow Label} is a 20-bit field intended to identify flows requiring special handling. Valid values lie in \([0,1{,}048{,}575]\). Larger values, bit-width misalignment, or incorrect formatting violate the specification and disrupt flow-based processing.

\subsection{DSCP Field Violation}
The \emph{Differentiated Services Code Point} (DSCP) occupies the first six bits of the Traffic Class field and must be in \([0,63]\). Values outside this range, or use of undefined/non-standard labels, constitute violations.

\subsection{ECN Field Violation}
The \emph{Explicit Congestion Notification} (ECN) field uses two bits and therefore permits values only in \([0,3]\). Out-of-range encodings suggest erroneous congestion signaling or intentional misconfiguration.

\subsection{Protocol Mismatch Violation}
The \emph{Next Header} value must correspond to the actual transport-layer payload (e.g., TCP, UDP, ICMPv6). Declaring one protocol while embedding another (e.g., \emph{Next Header} indicates TCP but the payload is UDP-like) creates parsing inconsistencies and is treated as non-compliance.

\subsection{Next Header Field Violation}
Valid \emph{Next Header} values include standard transport protocols (e.g., TCP = 6, UDP = 17, ICMPv6 = 58) and recognized extension headers. Unassigned, experimental, or nonsensical codes constitute violations of the IPv6 standard.

\subsection{ICMPv6 Code Violation}
ICMPv6 message types and codes must be drawn from the set defined by the standard (e.g., Echo Request, Neighbor Solicitation). Values outside the valid ranges are invalid, whether due to corruption or attempts to bypass protocol validation.

\section{Agent Prompts for AICCE Architectures}
\label{app:agent_prompts}

This appendix lists the prompts used by AICCE in (i) Explainability Mode (Architecture~A) and (ii) Script Execution Mode (Architecture~B).

\subsection{Architecture A: Explainability Mode (Section-Level Compliance + MAD)}
\label{app:prompts_archA}

\begin{formal}
\lstset{style=aiccecode}
\begin{lstlisting}
==========================
Section Compliance Agent (System Prompt)
==========================
You are an IPv6 standards expert for {SECTION_NAME}.

You will be given:
1) Curated general IPv6 rules (Rg)
2) A specific RFC section excerpt (Ci)
3) One IPv6 packet represented as a dictionary of observed fields/values

DECISION POLICY (critical):
- Use ONLY (a) explicit normative requirements in the provided section/excerpt and Rg,
  and (b) the packet fields present in the data sample.
- You MUST NOT mark the packet non-compliant because information is missing or unknown in that particular section.
- Output Non-compliance ONLY when an explicit MUST/SHALL/MUST NOT/SHALL NOT requirement is
  contradicted by an observed packet field value.
- Do NOT use SHOULD/RECOMMENDED/MAY as grounds for Non-compliance.
- Do NOT speculate beyond the provided packet fields, and do not request additional information.
- Keep the reasoning strictly tied to: (i) the normative clause and (ii) the packet fields used.

OUTPUT LABELS (critical):
- Your Line 1 MUST start with EXACTLY ONE of: "Yes", "No", "Maybe Yes", "Maybe No"
- "Yes"  = Compliant with respect to this section based on available fields.
- "No"   = Non-compliant with respect to this section due to a direct contradiction.
- "Maybe Yes" / "Maybe No" = borderline/uncertain due to interpretation tension; triggers MAD.

Trigger rule:
- Any output other than "Yes" (i.e., "No", "Maybe Yes", or "Maybe No") triggers debate.

General rules (Rg):
{GENERAL_RULES}


==========================
Section Compliance Agent (User Prompt)
==========================
Specific Section (Ci):
{SECTION_TEXT}

TASK:
Determine whether this packet is compliant with the provided section, under the decision policy.

IMPORTANT OUTPUT FORMAT:
Line 1: Yes OR No OR Maybe Yes OR Maybe No (exactly one of these strings)
Line 2: One sentence explaining (a) the concrete normative requirement and (b) the concrete packet field(s) used.

Data Sample:
{IPV6_SAMPLE_JSON}


==========================
Debate Agent (User Prompt)
==========================
Specific Section (Ci):
{SECTION_TEXT}

Previous agent conclusion: {PREV_VERDICT}
Previous agent reason:
{PREV_REASON}

DEBATE TASK:
Re-evaluate the packet strictly under the same decision policy.
- If the previous decision relied on missing information, non-normative language (SHOULD/MAY),
  or reasoning not grounded in the provided section/Rg + packet fields, you MUST correct it.
- If the previous decision missed a direct contradiction with a MUST/SHALL/MUST NOT/SHALL NOT clause,
  you MUST correct it.

IMPORTANT OUTPUT FORMAT:
Line 1: Yes OR No OR Maybe Yes OR Maybe No (exactly one of these strings)
Line 2: One sentence citing (a) the concrete normative requirement and (b) the concrete packet field(s) used.

Data Sample:
{IPV6_SAMPLE_JSON}
\end{lstlisting}
\end{formal}

\subsection{Architecture B: Script Execution Mode (Rule Extraction + Script Generation + Critic)}
\label{app:prompts_archB}

\begin{formal}
\lstset{style=aiccecode}
\begin{lstlisting}
==========================
Section Rule Extractor Agent (System Prompt)
==========================
You are an IPv6 standards compliance expert.
Your job is to extract checkable compliance rules.

Constraints:
- Only output rules whose required packet fields exist in the provided DataFrame profile.
- Output one rule per line, as a Python-like boolean expression.
- No explanations, no markdown, no numbering.
- Keep each rule concise and directly checkable.


==========================
Section Rule Extractor Agent (User Prompt)
==========================
SECTION NAME:
{SECTION_NAME}

SECTION TEXT:
{SECTION_TEXT}

GENERAL RULES (Rg):
{GENERAL_RULES}

DATAFRAME PROFILE (available columns/types/samples):
{DF_PROFILE}

TASK:
Extract all checkable compliance rules from the section + general rules, subject to the constraints.
Output one Python-like boolean condition per line.


==========================
Script Generator Agent (System Prompt for initial generation)
==========================
You are an expert Python developer and IPv6 standards expert.
You produce reliable, executable Python code.

Output constraints:
- Return ONLY Python code.
- Do NOT include markdown fences.
- Code MUST start with: def check_ipv6_compliance(df):
- Do NOT include explanatory text.
- Only implement rules that reference existing DataFrame columns.


==========================
Script Generator Agent (User Prompt for initial generation)
==========================
GENERAL RULES (Rg):
{GENERAL_RULES}

DATAFRAME PROFILE:
{DF_PROFILE}

EXTRACTED RULES (Python-like expressions, one per line):
{RULES_TEXT}

TASK:
Write a Python function named 'check_ipv6_compliance' that accepts a pandas.DataFrame df,
where each row is an IPv6 packet.

Return a DataFrame with:
- A 'Compliant' column (1 if all implemented rules pass on that row, else 0)
- One column per implemented rule (1 pass, 0 fail)

Implementation requirements:
- Safely handle missing values in existing columns (treat NaN as pass unless the rule explicitly
  requires otherwise).
- Never reference columns not present in df.
- Make rule column names stable (e.g., Rule_001, Rule_002, ...).
- Evaluate rules row-wise in a vectorized way when possible.


==========================
Script Generator Agent (System Prompt for retry mode)
==========================
You are an expert Python developer and IPv6 standards expert.
You must FIX broken code to run correctly.

Output constraints:
- Return ONLY Python code.
- Do NOT include markdown fences.
- Code MUST start with: def check_ipv6_compliance(df):
- Do NOT include explanatory text.
- Keep the logic aligned with provided rules and dataframe profile.


==========================
Script Generator Agent (User Prompt retry mode)
==========================
GENERAL RULES (Rg):
{GENERAL_RULES}

DATAFRAME PROFILE:
{DF_PROFILE}

EXTRACTED RULES:
{RULES_TEXT}

PREVIOUS CODE:
{PREV_SCRIPT}

ERROR MESSAGE:
{ERROR_MESSAGE}

TASK:
Fix the function so it executes successfully and follows the constraints.


==========================
Critic Agent (Script Reviewer) (System Prompt)
==========================
You are a strict Python code reviewer and IPv6 standards expert.
Review the code for correctness, completeness, and alignment with the rules.

Constraints:
- Only critique issues involving columns that exist in the provided DataFrame profile.
- If the script is correct and complete, output ONLY: APPROVED
- Otherwise, output specific fixes required.


==========================
Critic Agent (Script Reviewer) (User Prompt)
==========================
GENERAL RULES (Rg):
{GENERAL_RULES}

DATAFRAME PROFILE:
{DF_PROFILE}

RULES:
{RULES_TEXT}

SCRIPT:
{SCRIPT_CODE}
\end{lstlisting}
\end{formal}

\section{MAD Effect on Agent Reasoning}
\label{app:mad_effect_reasoning}

\subsection{Corrective Role of Multi-Agent Debate (MAD)}
\label{app:mad_positive_effect}

In most evaluated packets, Multi-Agent Debate (MAD) improves compliance classification by correcting single-pass errors and enforcing explicit reconciliation between retrieved clause evidence and cross-field packet structure. This effect is especially consistent for mid-tier and higher-capability LLMs, where MAD helps expose structural inconsistencies that can be missed when features are evaluated independently. In contrast, for weaker models, MAD can occasionally introduce additional uncertainty (e.g., unnecessary \texttt{Maybe} judgments) that leads to over-analysis and, in very rare cases, an incorrect final verdict.

\subsubsection{Case 1: TCP+UDP Overlap}
\label{app:mad_case1_tcp_udp}

To illustrate this corrective behavior, we consider a challenging overlap scenario in which a packet record simultaneously contains populated TCP and UDP-specific header fields. Such TCP/UDP overlap cases are difficult for single-pass LLM reasoning because each TCP or UDP-specific field may appear individually well-formed and within its own compliance constraints, yet the \emph{joint} presence of both transport headers violates transport-layer coherence. Under IPv6, a packet’s transport header must be consistent with \texttt{Protocol}/\texttt{Next Header}, and TCP and UDP headers cannot both apply to the same packet; therefore, any record exhibiting concurrent TCP and UDP header attributes should be classified as \emph{Non-compliant}.

Consider this Packet input:
\begin{formal}
\lstset{style=aiccecode}
\begin{lstlisting}
{
  "Time": "00:00:01.836824",
  "Host": "2a43:c071:ea8e:e12e:aef9:a3f7:ac:68e4,2865:8e9:f7fb:7ef3:ca00:f264:f33f:f836",
  "Source": "2a43:c071:ea8e:e12e:aef9:a3f7:ac:68e4",
  "Destination": "2865:8e9:f7fb:7ef3:ca00:f264:f33f:f836",
  "Protocol": "TCP",
  "Length": 76,
  "Info": "33959  >  443 [ACK] Seq=1 Ack=1 Win=113 Len=0[Packet size limited during capture]",
  "Hop Limit": 54,
  "Frag Header": null,
  "Routing Header": null,
  "DSCP": "Default",
  "ECN": "Not ECN-Capable Transport",
  "Freq/Channel": null,
  "Pad1": null,
  "PadN": null,
  "IPv6 Version": 6,
  "FragCount": null,
  "FlowLabel": "0x00000",
  "Bogus_Version": null,
  "Hop-By-Hop": null,
  "Frag_Error": null,
  "Frag_Overlap": null,
  "Next Header": "TCP",
  "Source Port Resolved": 33959.0,
  "Source Port Unresolved": 33959.0,
  "Destination Port Resolved": 443.0,
  "Destination Port Unresolved": 443.0,

  "TCP Src Port": 33959.0,
  "TCP Dst Port": 443.0,
  "TCP Stream": 86593.0,
  "TCP Sequence": 1.0,
  "TCP Sequence Raw": 2110441923.0,
  "TCP Acknowledgement": 1.0,
  "TCP Acknowledgement Raw": 3830669244.0,
  "TCP Header Length": 32.0,
  "TCP Flags": "0x010",
  "TCP Window Size": 113.0,
  "TCP Checksum": "0xd0a4",
  "TCP Options": "0101080a",

  "UDP Source Port": 53.0,
  "UDP Destination Port": 44551.0,
  "UDP Stream": 271081.0,
  "UDP Length": 98.0,
  "UDP Checksum": "0x8fbf",

  "ICMPv6 Type": null,
  "ICMPv6 Code": null,
  "ICMPv6 Checksum": null,
  "ICMPv6 Length": null,
  "ICMPv6 Data": null
}
\end{lstlisting}
\end{formal}

\paragraph{Explainability Mode (without MAD)}
Using a representative mid-tier model (\texttt{Llama-3.1-8B}), single-pass evaluation can overemphasize per-field plausibility (e.g., valid port ranges, checksums, and header values) and fail to enforce transport-layer consistency across fields. As a result, all agents return a compliant verdict and the system outputs an incorrect final decision.

\begin{nomad}
\lstset{style=aiccecode}
\begin{lstlisting}
[NO-MAD] Model=Llama-3.1-8B | Final Verdict=Compliant

compliance_agent1  => Compliant
compliance_agent2  => Compliant
compliance_agent3  => Compliant
compliance_agent4  => Compliant
compliance_agent5  => Compliant
compliance_agent6  => Compliant
compliance_agent7  => Compliant
compliance_agent8  => Compliant
compliance_agent9  => Compliant
compliance_agent10 => Compliant
compliance_agent11 => Compliant
compliance_agent12 => Compliant
compliance_agent13 => Compliant
compliance_agent14 => Compliant
compliance_agent15 => Compliant
compliance_agent16 => Compliant

[FINAL] COMPLIANT
\end{lstlisting}
\end{nomad}

\paragraph{Explainability Mode (with MAD)}
With MAD enabled, uncertainty in a subset of section-level judgments triggers a debate phase. During debate, agents explicitly validate the coherence between \texttt{Protocol}/\texttt{Next Header} and the presence of transport-specific fields, and they identify that TCP and UDP headers cannot simultaneously apply to one packet record. This yields the correct \emph{Non-compliant} verdict.

\begin{mad}
\lstset{style=aiccecode}
\begin{lstlisting}
[MAD-ENABLED] Model=Llama-3.1-8B | Final Verdict=Non-compliant | Debate Rounds=2

--- Round 0: Initial Section Judgments (Yes/Maybe/No) ---
compliance_agent1  => Yes
compliance_agent2  => Yes
compliance_agent3  => Yes
compliance_agent4  => Yes
compliance_agent5  => Yes
compliance_agent6  => Yes
compliance_agent7  => Yes
compliance_agent8  => Yes
compliance_agent9  => Maybe No
compliance_agent10 => Yes
compliance_agent11 => Yes
compliance_agent12 => Yes
compliance_agent13 => Yes
compliance_agent14 => Yes
compliance_agent15 => Yes
compliance_agent16 => No

--- Debate Round 1 ---
[DEBATE 1] compliance_agent9  => No
[REASON] Protocol/Next Header indicate TCP, yet UDP Source/Destination Port, UDP Length, and UDP Checksum are populated. A packet cannot carry both TCP and UDP transport headers; this is a transport-structure inconsistency.

[DEBATE 1] compliance_agent16 => No
[REASON] When TCP is the transport header, UDP fields should be absent (and vice versa).

--- Debate Round 2 ---
[DEBATE 2] compliance_agent9  => No
[REASON] Debate Agent agrees with prior assessment. The simultaneous presence of TCP and UDP header fields violates transport-header coherence.

--- Final Aggregation ---
compliance_agent1  => Compliant
compliance_agent2  => Compliant
compliance_agent3  => Compliant
compliance_agent4  => Compliant
compliance_agent5  => Compliant
compliance_agent6  => Compliant
compliance_agent7  => Compliant
compliance_agent8  => Compliant
compliance_agent9  => Non-compliant
compliance_agent10 => Compliant
compliance_agent11 => Compliant
compliance_agent12 => Compliant
compliance_agent13 => Compliant
compliance_agent14 => Compliant
compliance_agent15 => Compliant
compliance_agent16 => Non-compliant

[FINAL] NON-COMPLIANT
[REASON] (system-level): MAD identifies a TCP/UDP overlap that violates transport-header coherence; any Non-compliant section outcome yields a Non-compliant final verdict.
\end{lstlisting}
\end{mad}

\subsubsection{Case 2: Flow Label Set to Zero}
\label{app:mad_case2_flowlabel_zero}

While MAD is designed to be corrective, it is not theoretically guaranteed to improve every decision for every input. In rare edge cases, debate may be triggered by conservative uncertainty (e.g., an initial \texttt{Maybe Yes} judgment) on an otherwise compliant packet, and subsequent agents may converge on an incorrect interpretation that is not grounded in the clause semantics relevant to the current packet context. Although such negative impacts are extremely rare in our evaluations, the example below illustrates this failure mode using a compliant packet where the IPv6 \texttt{FlowLabel} is set to \texttt{0x00000}. In real traffic, \texttt{FlowLabel=0} is valid and commonly observed; therefore, the correct verdict for this packet is \emph{Compliant}.

Consider this Packet input:
\begin{formal}
\lstset{style=aiccecode}
\begin{lstlisting}
{
  "Time": "00:00:16.943454",
  "Host": "2a43:10fe:b717:7fa1:3878:ebad:c72e:f8bc,2865:fb:e77c:2218:be91:73a2:e730:3575",
  "Source": "2a43:10fe:b717:7fa1:3878:ebad:c72e:f8bc",
  "Destination": "2865:fb:e77c:2218:be91:73a2:e730:3575",
  "Protocol": "ICMPv6",
  "Length": 84,
  "Info": "Echo (ping) reply id=0x62da, seq=10, hop limit=54",
  "Hop Limit": 54,
  "Frag Header": null,
  "Routing Header": null,
  "DSCP": "Default",
  "ECN": "Not ECN-Capable Transport",
  "Freq/Channel": null,
  "Pad1": null,
  "PadN": null,
  "IPv6 Version": 6,
  "FragCount": null,
  "FlowLabel": "0x00000",
  "Bogus_Version": null,
  "Hop-By-Hop": null,
  "Frag_Error": null,
  "Frag_Overlap": null,
  "Next Header": "ICMPv6",

  "Source Port Resolved": null,
  "Source Port Unresolved": null,
  "Destination Port Resolved": null,
  "Destination Port Unresolved": null,

  "TCP Src Port": null,
  "TCP Dst Port": null,
  "TCP Stream": null,
  "TCP Sequence": null,
  "TCP Sequence Raw": null,
  "TCP Acknowledgement": null,
  "TCP Acknowledgement Raw": null,
  "TCP Header Length": null,
  "TCP Flags": null,
  "TCP Window Size": null,
  "TCP Checksum": null,
  "TCP Options": null,

  "UDP Source Port": null,
  "UDP Destination Port": null,
  "UDP Stream": null,
  "UDP Length": null,
  "UDP Checksum": null,

  "ICMPv6 Type": "Echo (ping) reply",
  "ICMPv6 Code": 0,
  "ICMPv6 Checksum": "0x1389",
  "ICMPv6 Length": null,
  "ICMPv6 Data": null
}
\end{lstlisting}
\end{formal}

\paragraph{Explainability Mode (without MAD)}
In a single-pass setting, the model may correctly treat \texttt{FlowLabel=0} as valid and produce the correct \emph{Compliant} decision as below.

\begin{nomad}
\lstset{style=aiccecode}
\begin{lstlisting}
[NO-MAD] Model=Llama-3.1-8B | Final Verdict=Compliant

compliance_agent1  => Compliant
compliance_agent2  => Compliant
compliance_agent3  => Compliant
compliance_agent4  => Compliant
compliance_agent5  => Compliant
compliance_agent6  => Compliant
compliance_agent7  => Compliant
compliance_agent8  => Compliant
compliance_agent9  => Compliant
compliance_agent10 => Compliant
compliance_agent11 => Compliant
compliance_agent12 => Compliant
compliance_agent13 => Compliant
compliance_agent14 => Compliant
compliance_agent15 => Compliant
compliance_agent16 => Compliant

[FINAL] COMPLIANT
\end{lstlisting}
\end{nomad}

\paragraph{Explainability Mode (with MAD)}
With MAD enabled, an early \texttt{Maybe Yes} judgment can trigger debate even though the packet is compliant. In this example, subsequent agents incorrectly enforce a non-existent constraint (e.g., that a flow label must be non-zero or randomized) and converge to an incorrect \emph{Non-compliant} verdict. This illustrates that MAD can occasionally over-correct when debate is triggered on a borderline interpretation and later rounds amplify an incorrect assumption, although it is a rare occurrence.

\begin{mad}
\lstset{style=aiccecode}
\begin{lstlisting}
[MAD-ENABLED] Model=Llama-3.1-8B | Final Verdict=Non-compliant | Debate Rounds=2

--- Round 0: Initial Section Judgments (Yes/Maybe/No) ---
compliance_agent1  => Yes
compliance_agent2  => Yes
compliance_agent3  => Yes
compliance_agent4  => Yes
compliance_agent5  => Yes
compliance_agent6  => Maybe Yes
compliance_agent7  => Yes
compliance_agent8  => Yes
compliance_agent9  => Yes
compliance_agent10 => Yes
compliance_agent11 => Yes
compliance_agent12 => Yes
compliance_agent13 => Yes
compliance_agent14 => Yes
compliance_agent15 => Yes
compliance_agent16 => Yes

--- Debate Round 1 ---
[DEBATE 1] compliance_agent6  => Maybe No
[REASON] Should flag uncertainty about FlowLabel=0, requesting additional verification.

--- Debate Round 2 ---
[DEBATE 2] compliance_agent6 => No
[Reason] FlowLabel 0 appears to be an unique case. Debate phase interprets such case as suspicious and non-compliant.

--- Final Aggregation ---
compliance_agent1  => Compliant
compliance_agent2  => Compliant
compliance_agent3  => Compliant
compliance_agent4  => Compliant
compliance_agent5  => Compliant
compliance_agent6  => Non-compliant
compliance_agent7  => Compliant
compliance_agent8  => Compliant
compliance_agent9  => Compliant
compliance_agent10 => Compliant
compliance_agent11 => Compliant
compliance_agent12 => Compliant
compliance_agent13 => Compliant
compliance_agent14 => Compliant
compliance_agent15 => Compliant
compliance_agent16 => Compliant

[FINAL] NON-COMPLIANT
[REASON] (system-level): Debate converges on a FlowLabel constraint; any Non-compliant section outcome yields a Non-compliant final verdict.

\end{lstlisting}
\end{mad}

\end{document}